\documentclass[preprint,
superscriptaddress,
 amsmath,amssymb,
 aps,
 prl
]{revtex4-2}
\DeclareUnicodeCharacter{2212}{-}
\usepackage{graphicx}
\usepackage{dcolumn}
\usepackage{bm}
\usepackage{units}
\usepackage{float}
\usepackage{amsmath}
\renewcommand{\thesection}{\Roman{section}} 
\usepackage{dblfloatfix}
\usepackage{xcolor}
\usepackage[utf8]{inputenc}
\usepackage[T1]{fontenc}

\usepackage{amsfonts}
\usepackage{booktabs}

\usepackage{hhline}
\usepackage{multirow}
\usepackage{array}

\newcolumntype{P}[1]{>{\centering\arraybackslash}p{#1}}

\usepackage{mathtools}

\begin{document}

\title{Single-pulse Gy-scale irradiation of biological cells at $10^{13}$ Gy/s average dose-rates from a laser-wakefield accelerator}

\author{C.A.~McAnespie}
\affiliation{School of Mathematics and Physics,
  Queen's University Belfast, BT7 1NN, Belfast United Kingdom}

  \author{P. Chaudhary} 
\affiliation{Radiotherapy and Dosimetry Group
National Physical Laboratory
Hampton Road, Teddington, Middlesex
TW11 0LW, UK}

   \author{M.J.V.~Streeter} 
\affiliation{School of Mathematics and Physics,
  Queen's University Belfast, BT7 1NN, Belfast United Kingdom}

  \author{S.W.~Botchway} 
\affiliation{Central Laser Facility, Rutherford Appleton Laboratory, Harwell Campus, Didcot, OX11 0QX, Oxford, UK}

  \author{N. Bourgeois} 
\affiliation{Central Laser Facility, Rutherford Appleton Laboratory, Harwell Campus, Didcot, OX11 0QX, Oxford, UK}
 
  \author{L. Calvin}
\affiliation{School of Mathematics and Physics,
  Queen's University Belfast, BT7 1NN, Belfast United Kingdom}

  \author{N. Cavanagh}
\affiliation{School of Mathematics and Physics,
  Queen's University Belfast, BT7 1NN, Belfast United Kingdom}

   \author{K. Fleck}
\affiliation{School of Mathematics and Physics,
  Queen's University Belfast, BT7 1NN, Belfast United Kingdom}

   \author{D. Jaroszynski} 
\affiliation{SUPA, Department of Physics, University of Strathclyde, Glasgow, G4 0NG, UK}

   \author{B. Kettle} 
\affiliation{Imperial College London, London, SW7 2AZ, UK}

   \author{A.M.~Lupu} 
\affiliation{Extreme Light Infrastructure-Nuclear Physics, Magurele, Romania}

   \author{S.P.D.~Mangles} 
\affiliation{Imperial College London, London, SW7 2AZ, UK}

 \author{S. J.~McMahon} 
\affiliation{Patrick G. Johnston Centre for Cancer Research, 
  Queen's University Belfast, BT7 1NN, Belfast United Kingdom}

    \author{J. Mill} 
\affiliation{SUPA, Department of Physics, University of Strathclyde, Glasgow, G4 0NG, UK}

    \author{S. R.~Needham} 
\affiliation{Central Laser Facility, Rutherford Appleton Laboratory, Harwell Campus, Didcot, OX11 0QX, Oxford, UK}

   \author{P. P.~Rajeev} 
\affiliation{Central Laser Facility, Rutherford Appleton Laboratory, Harwell Campus, Didcot, OX11 0QX, Oxford, UK}

\author{J. Sarma}
\affiliation{School of Mathematics and Physics,
  Queen's University Belfast, BT7 1NN, Belfast United Kingdom}

  \author{K. M. Prise} 
\affiliation{Patrick G. Johnston Centre for Cancer Research, Queen's University Belfast, BT7 1NN, Belfast United Kingdom}
 
 \author{G. Sarri} 
\affiliation{School of Mathematics and Physics,
  Queen's University Belfast, BT7 1NN, Belfast United Kingdom}

\begin{abstract}
We report on the first experimental characterization of a laser-wakefield accelerator able to deliver, in a single pulse, doses in excess of \unit[1]{Gy} on timescales of the order of a hundred femtoseconds, reaching unprecedented average dose-rates up to \unit[10$^{13}$]{Gy/s}. The irradiator is demonstrated to deliver doses tuneable up to \unit[2.2]{Gy} in a cm$^2$ area and with a high degree of longitudinal and transverse uniformity in a single irradiation. In this regime, proof-of-principle irradiation of patient-derived glioblastoma stem-like cells and human skin fibroblast cells show indications of a differential cellular response, when compared to reference irradiations at conventional dose-rates. These include a statistically significant increase in relative biological effectiveness ($1.40\pm0.08$ at 50\% survival for both cell lines) and a significant reduction of the relative radioresistance of tumour cells. Data analysis provides preliminary indications that these effects might not be fully explained by induced oxygen depletion in the cells but may be instead linked to a higher complexity of the damages triggered by the ultra-high density of ionising tracks of femtosecond-scale radiation pulses. These results demonstrate an integrated platform for systematic radiobiological studies at unprecedented beam durations and dose-rates, a unique infrastructure for translational research in radiobiology at the femtosecond scale.

\end{abstract}

\maketitle

\newpage

\section*{Introduction}
Radiotherapeutic methods are among the most effective ways of treating cancer \cite{ahmad2012advances}, even though they are known to be prone to significant side effects, such as the development of secondary cancers \cite{American} and tissue damage that may require surgical intervention \cite{salvo2010prophylaxis}. The potentially irreversible and harmful secondary effects of radiotherapy have resulted in a generally conservative approach by practitioners. This approach is fully justified, if one considers that a fundamental understanding of cellular response to ionising radiation requires disentangling an intricate web of processes and mechanisms, crossing boundaries between physics, chemistry, and biology. These processes typically span more than twenty orders of magnitude in time, starting with ionisation and the generation and diffusion of radicals (from a few to hundreds of femtoseconds after irradiation), through the chemical reactions inducing DNA damage and repair (from picoseconds up to minutes after irradiation), and ending with the macroscopic effects induced by the biological response of the cells (emerging even years after irradiation) \cite{adams1980time}. 

The primary role of ionising radiation is to produce radicals in the cell, a task that is typically completed within the first few to hundreds of femtoseconds after entering biological tissue \cite{adams1980time}. The water radical cation H$_2$O$^{\cdot +}$ then undergoes an ultrafast proton transfer reaction to a neighbouring water molecule to form the hydroxyl radical \cite{loh2020observation}, while the electrons solvate within 200 – \unit[500]{fs} \cite{Low2022}. Hydroxyl radicals can trigger extensive oxidative damage in the cell, while solvated electrons may cause genomic damage by dissociative electron attachment \cite{alizadeh2012precursors}. The potential effects of continued dose delivery during these early physico-chemical stages is largely unknown but is expected to be non-negligible. It is possible that protracted irradiation times might significantly impact the efficiency of signalling channels and repair processes, and on the recruitment of transducer proteins for programmed cell death or senescence \cite{harrington2019ultrahigh}. As an example of this secondary effect of radiation, it is known that radiation delivered after the solvation of the electron ($>$\unit[500]{fs}) can induce its excitation, thus altering its reactivity \cite{Low2022}. 

Conventional radiotherapy typically delivers Gy-scale doses over a range of minutes, using trains of $\mu$s-long mGy pulses \cite{schuler2022ultra}. Also the FLASH-RT regime, while achieving relatively higher dose-rates in the range of 10s – \unit[100s]{Gy/s} \cite{harrington2019ultrahigh,Leavitt}, still operates on timescales that are significantly longer than the cell’s physical response \cite{adams1980time}.
In order to minimise secondary effects triggered by protracted irradiation, it is instead essential to enter a fundamentally different regime, where Gy-scale doses are delivered in a single pulse with a duration comparable to that of the physical response of biological tissue (i.e., $\lesssim$ \unit[100]{fs}), resulting in average dose-rates above \unit[10$^{12}$]{Gy/s}.
Radiation delivery at such a high dose-rate will also be associated with a much higher density of ionising tracks and radical clouds, which might induce non-linear effects in the radical chemistry and in the processes involved in DNA damage and repair. 
While recent numerical work suggests that these effects might be negligible up to the nanosecond timescale \cite{Thompson} and no experimental evidence has been thus far observed on differential response down to tens of picoseconds (see, e.g., Refs. \cite{hill2002increased,mcanespie2023ijrobp}), the importance of these effects is still largely unknown for femtosecond-scale radiation. 
The ultra-high dose-rate (UHDR) regime accessed by femtosecond-scale radiation is thus fundamentally distinct from FLASH-RT deliveries \cite{harrington2019ultrahigh,Leavitt} and the observed potential benefits of FLASH-RT irradiation are likely to be triggered by mechanisms of a different nature.

\begin{figure}[t]
\centering
\includegraphics[width=\linewidth]{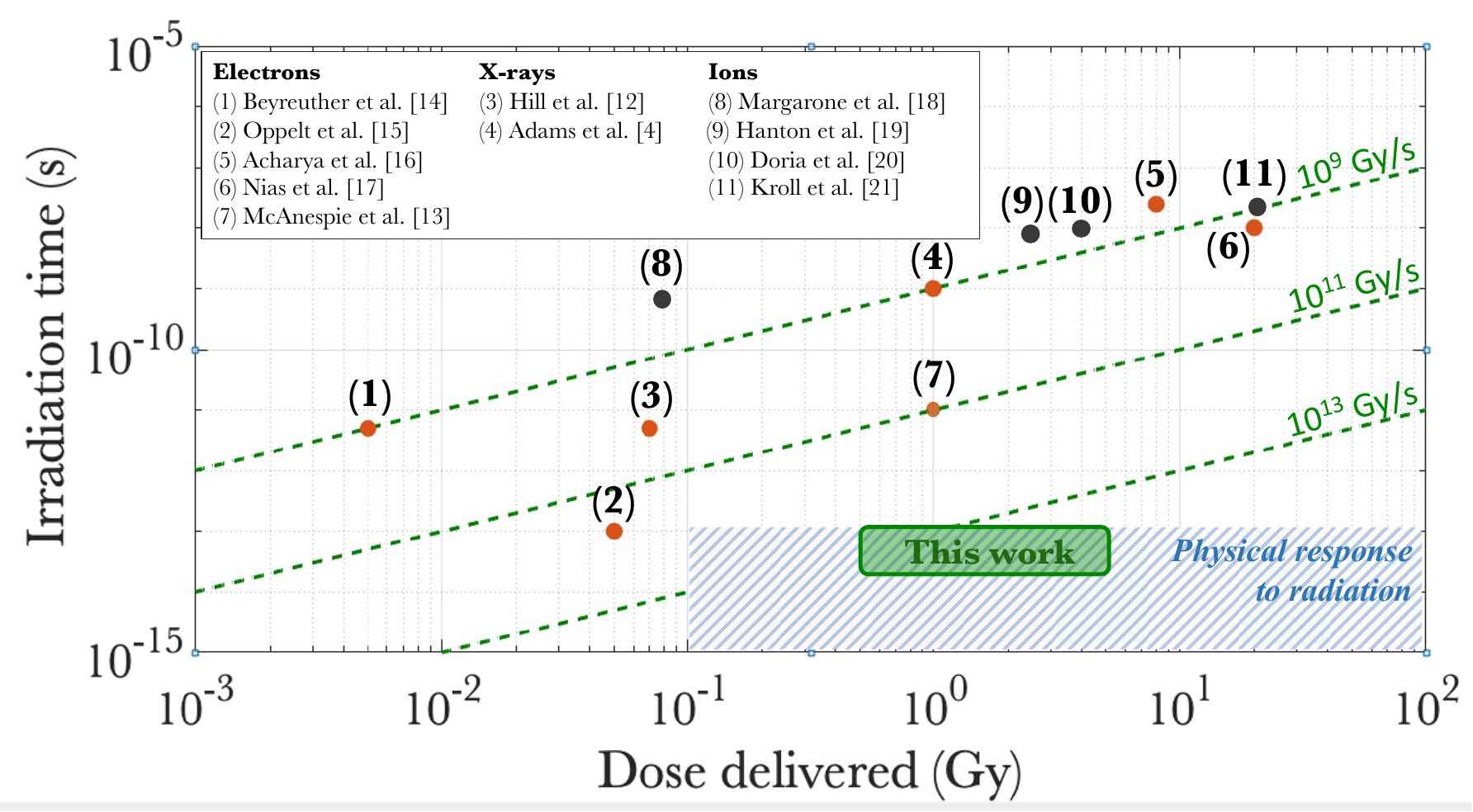}
\caption{\textbf{Timescales accessed in this work compared to state of the art:} Single-shot dose and irradiation times accessed in this study (green rectangle) compared with the typical timescale of the physical response to radiation (blue band) and representative experimental work reported in the literature for electron, photon (orange dots) and ion (grey dots) irradiations. Dose-rate isocurves are shown for comparison (green dashed lines).}
\label{fig1}
\end{figure}

Laser-driven particle accelerators have been identified as potential tools to access this novel regime of radiobiology, due to the inherently short duration of the particle beams that they can produce. Proof-of-principle experimental studies of the biological effect of UHDR radiation have been recently reported using a wide range of radiation and particle beams (see, e.g., Refs. \cite{adams1980time, beyreuther2015radiobiological, oppelt2015comparison, mcanespie2023ijrobp, hill2002increased, acharya2011dose,Nias,margarone2018,hanton2019dna,doria2012biological,kroll_2022}, and Fig.\ref{fig1}). However, these could only achieve Gy-scale
irradiations either via dose fractionation \cite{oppelt2015comparison,beyreuther2015radiobiological} (thus still resulting in average delivery times of the order of minutes) or on timescales of the order of nanoseconds \cite{adams1980time, acharya2011dose,Nias,margarone2018,hanton2019dna,doria2012biological,kroll_2022} or tens of picoseconds \cite{hill2002increased, mcanespie2023ijrobp}. In all cases, no definitive deviation of radiobiological response from dose delivery at conventional dose-rates has been reported. 

Here, we experimentally demonstrate that single-pulse delivery of Gy-scale doses at the femtosecond level can now be achieved using a wakefield accelerator (LWFA, see, e.g., Ref. \cite{esarey2009physics}) driven by oversized laser focal spots \cite{poder_2024}. 
The capability of performing single-shot irradiations, together with the on-shot full characterisation of the dose deposited, makes the experimental platform robust to shot-to-shot variations in the electron beam characteristics. The wakefield accelerator, driven by a 100 TW-class laser system, is shown to deliver doses tuneable from zero up to \unit[2.2]{Gy} at a duration of the order of \unit[150]{fs}, resulting in average dose-rates up to \unit[10$^{13}$]{Gy/s}. 
This novel accelerator configuration provides a step-change enhancement in capability when compared to other laser-driven sources able to deliver Gy-scale doses at \unit[20]{ps} \cite{mcanespie2023ijrobp,McAnespie_pre}, with a reduction in dose delivery times of more than two orders of magnitude that allows to access the physical stage of cellular response to radiation.
The viability of this radiation source for radiobiological studies at the ultra-high dose-rate frontier has been demonstrated by performing pilot irradiations of patient-derived glioblastoma stem-like cells (E2 cell line) and human skin fibroblast cells (AG01522 D cell line), typical models to mimic tumour and healthy tissue in the laboratory. 
The results have been compared to irradiation at a conventional dose-rate of \unit[0.49]{Gy/min} using a \unit[225]{kVp} x-ray source, a widely accepted standard in comparative radiobiology against which relative biological effectiveness (RBE) is typically calculated (see, e.g., Ref. \cite{Murshed}). Even though the energy of the photons is much lower than that of the electrons from a wakefield accelerator, we note that recent numerical work has demonstrated that doses delivered by high energy electron beams at conventional dose-rates do not induce a statistically significant change in biological outcome, with an RBE still approximately equal to 1 \cite{Delorme} (consistently with proof-of-principle experiments \cite{beyreuther2015radiobiological, oppelt2015comparison}), justifying a direct comparison between the x-ray data and the femtosecond-scale irradiation reported here.

When compared to reference irradiations at conventional dose-rates, clonogenic survival assays and DNA damage measurements by 53BP1 foci induction under normoxic and hypoxic conditions provided preliminary indication of differential response to femtosecond-scale radiation. 
Both tumour and normal cell lines exhibited an RBE = $1.40\pm0.08$ at 50\% survival, in conjunction with a higher concentration of damage sites per cell with a statistically significant larger foci size. 
Interestingly, the typically higher radioresistance of tumour cells was also considerably reduced at these ultra-high dose rates. Recent results obtained at picosecond timescales ($\approx$10 - \unit[20]{ps}) did not provide clear evidence of these effects  \cite{hill2002increased, mcanespie2023ijrobp}, suggesting that they might be related to novel effects arising at these ultra-short timescales. 
This experimental platform provides a novel and integrated infrastructure for radiobiological research at the femtosecond scale, i.e., at timescales comparable to the physical response of cells to radiation \cite{adams1980time}.
The demonstrated irradiation area and spatial uniformity are consistent with those used in other proof-of-principle in-vivo irradiation (see, e.g., Ref. \cite{kroll_2022}). The proof-of-principle demonstrator presented here thus provides hypotheses-generating evidence that lays the foundations for further steps including in-vivo and pre-clinical studies.

\section*{Experimental setup and electron beam properties}

\begin{figure*}[t]
\centering
\includegraphics[width=1\linewidth]{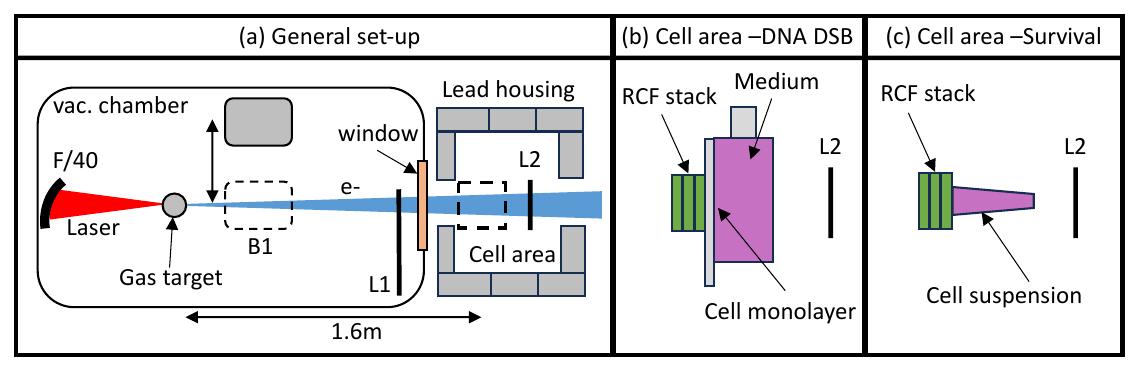}
\caption{\textbf{Sketch of the experimental set-up}:\textbf{(a)} Top-view sketch of the experimental setup. L1 and L2 are calibrated Lanex scinitillator screens used for spectrum and dose measurements, respectively. B1 is a removable \unit[30]{cm}-long \unit[1]{T} dipole magnet. The laser-driven accelerator was placed in a vacuum chamber with the electrons propagating, through a thin vacuum window made of Mylar and Kapton, onto a shielded cell irradiation area in air. Sketch of the cell irradiation area used for \textbf{(b)} the foci formation studies and \textbf{(c)} for the clonogenic assays.}
\label{fig2}
\end{figure*}

The experiment (sketched in Fig. \ref{fig2}) was carried out using the Gemini laser at the Central Laser Facility, UK, which delivered laser pulses with \unit[(6.7 $\pm$ 0.3)]{J} of encircled energy in focus in \unit[(45 $\pm$ 5)]{fs}.
The laser was focused using an F/40 off-axis parabola \unit[5]{mm} above a \unit[15]{mm} long supersonic gas-jet to drive the laser wakefield accelerator, in conditions similar to Ref. \cite{poder_2024}.
A \unit[98]{\%} He - \unit[2]{\%} N$_2$ gas mixture was used, and the electron bunches were produced following ionisation injection \cite{pak2010injection, mirzaie2015demonstration}, resulting in broadband and high-charge ultra-relativistic beams.
The electron plasma density in the accelerator was measured via means of optical interferometry to be $n_e$ = \unit[($4.0\pm0.4$)$\times$10$^{18}$]{cm$^{-3}$}.
The residual laser exiting the gas-jet was dumped by a \unit[1]{mm} ceramic screen placed \unit[1]{m} downstream of the gas-jet.

In this article, we present results of foci formation and clonogenic assays, which require different electron beam characteristics. Foci formation studies generally require a larger irradiated area with a deposited dose of the order of \unit[1]{Gy}, while clonogenic assays require higher and tuneable peak doses to be delivered in a smaller region. 
After an extensive parametric scan of the performance of the accelerator, it was found that the electron beam divergence - and, therefore, the beam size at the irradiation plane - could be controlled by varying the laser intensity in focus, with smaller beam divergences obtained at higher laser intensities. 
This can be intuitively understood by considering that, in this regime, a higher laser intensity results in stronger electron cavitation, which in turn may produce higher focussing fields in the accelerating cavity \cite{esarey2009physics,lu_2007}.
The laser intensity in focus was controlled by modifying the size of the laser focal spot using an adaptive optic placed before the off-axis parabola.
For the foci formation studies, the laser was focussed down to a focal spot size of \unit[(52.4 $\pm$ 1.8)]{\textmu m} by \unit[(72.8 $\pm$ 1.8)]{\textmu m}, resulting in a peak intensity of $(3.9\pm0.4)\times10^{18}$ Wcm$^{-2}$ (dimensionless laser intensity $a_0 = 1.4\pm0.1$).
For the clonogenic assays, the laser was focussed down to a focal spot size of \unit[(46.3 $\pm$ 0.9)]{\textmu m} by \unit[(60.6 $\pm$ 1.7)]{\textmu m}, resulting in a peak intensity of $(5.4\pm0.5)\times10^{18}$ Wcm$^{-2}$ (dimensionless laser intensity $a_0 = 1.6\pm0.1$).

\begin{figure}[H]
\centering
\includegraphics[width=1\linewidth]{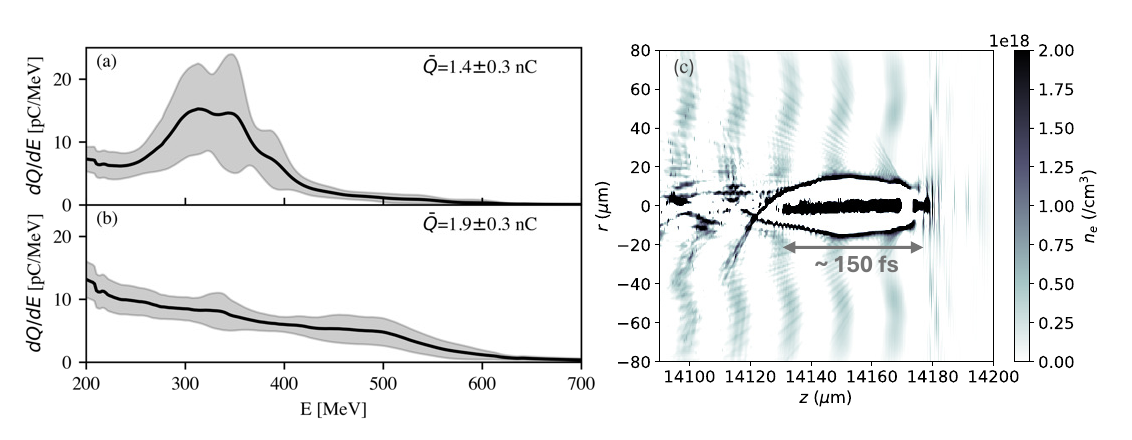}
\caption{\textbf{Electron beam properties}: Measured averaged angularly integrated spectra of the electron beams used for the \textbf{(a)} foci formation studies and \textbf{(b)} clonogenic assays. Black solid lines and grey shaded areas indicate the mean electron spectrum and the associated standard deviation, respectively, over an average of 30 shots. The mean and standard deviation of total charge in the beam is shown as $\Bar{Q}$. \textbf{(c)} Snapshot of the electron density at the end of the electron beam acceleration from a matching FB-PIC simulation (details in the text), showing the accelerating cavity and an electron beam duration of the order of \unit[150]{fs}.}
\label{fig3}
\end{figure}

The electron beam from LWFA was diagnosed, immediately before and after each cell irradiation run, using a magnetic spectrometer consisting of a \unit[30]{cm}, \unit[1]{T} dipole magnet and a LANEX scintillator screen \cite{glinec2006absolute}, allowing for the simultaneous measurement of spectrum, energy-dependent divergence, and total charge of the beam. 
The measured electron spectra used for the foci formation studies and the clonogenic assays are shown in Figs. \ref{fig3}.a and \ref{fig3}.b, together with their standard deviation. The divergence of the electron beam was measured at the cell plane to be \unit[$3.6\pm1.1$]{mrad} and \unit[$2.3\pm0.3$]{mrad} for the 
foci formation and the survival studies, respectively.

The electron beams presented relatively flat spectra extending from \unit[200]{MeV} (lower detection limit of the magnetic spectrometer) up to \unit[600]{MeV}, as typically expected from LWFA following ionisation injection \cite{pak2010injection, mirzaie2015demonstration}, and contained a total charge in the range of 1 - \unit[2]{nC}. 
The different spectral shape of the electron beams can be qualitatively understood by considering that, in the initial phases of the wakefield acceleration process, the laser beam undergoes strong self-focusing, which is a highly non-linear process dependent on the incident intensity distribution. This can significantly affect the injection and acceleration processes \cite{poder_2024,Mangles_2012} and, thus, the spectral shape and total charge of the electron beam. While the electron beam spectrum could only be measured immediately before and after irradiation, we note that spectral variations of this kind have a negligible effect on the dose deposited (e.g., variation of only 1\% between electrons at 200 and 600 MeV). The deposited dose is thus predominantly dictated by the electron beam charge and spatial distribution at the sample.

Since it was not possible to measure the electron beam duration directly, it has been inferred from matching Particle-In-Cell simulations using the FBPPIC code \cite{FBPIC}. 
The simulation assumes a \unit[15]{mm} plasma with an electron density of \unit[4.0$\times$10$^{18}$]{cm$^{-3}$} composed of 98\% of helium and 2\% of nitrogen with \unit[1]{mm} cosine ramps at either side. 
The simulation box lengths were \unit[145]{\textmu m} and \unit[90]{\textmu m} along the longitudinal (z) and radial (r) directions respectively, with a resolution of  \unit[50]{nm} $\times$ \unit[0.8]{\textmu m}. 
A moving window set-up was used to simulate the entire plasma length. The particles per cell were 2 along z, 2 along r and 8 along $\theta$, and two azimuthal modes were used. 
A Gaussian laser pulse with a central wavelength of \unit[800]{nm} and a pulse length of \unit[45]{fs} at FWHM was focused inside the plasma to a spot size of \unit[53]{\textmu m} corresponding to a normalised vector potential $a_0$=1.6. 
A snapshot of the electron density in the r-z plane at the end of the acceleration inside the plasma is shown in Fig. \ref{fig3}.c. 
The accelerated electron beam presents a similar spectrum and overall charge to those experimentally measured. 
The accelerating bubble extends for approximately three plasma periods and the electron beam has an overall duration of the order of \unit[150]{fs}, which will be assumed hereafter.


\section*{Dose properties at the cell irradiation plane}
The dipole magnet was removed for each irradiation run, with the electron beam propagating into a shielded cell-irradiation area (see Fig.\ref{fig2}). The cell cultures were placed \unit[1.7]{m} downstream of the electron beam source.
Here, a second LANEX screen and absolutely calibrated EBT3 Gafchromic films \cite{marroquin2016evaluation} were used to monitor, on-shot, the beam transverse profile and the spatial distribution of dose deposited (see Fig. 2). The procedure to absolutely calibrate the RCF films is discussed in Ref. \cite{McAnespie_pre}. The LANEX screen was calibrated by cross-comparison with the RCF films, with the dose found to be linear with scintillation counts up to a maximum of \unit[4]{Gy} \cite{McAnespie_pre}.

\begin{figure}[H]
\centering
\includegraphics[width=0.5\linewidth]{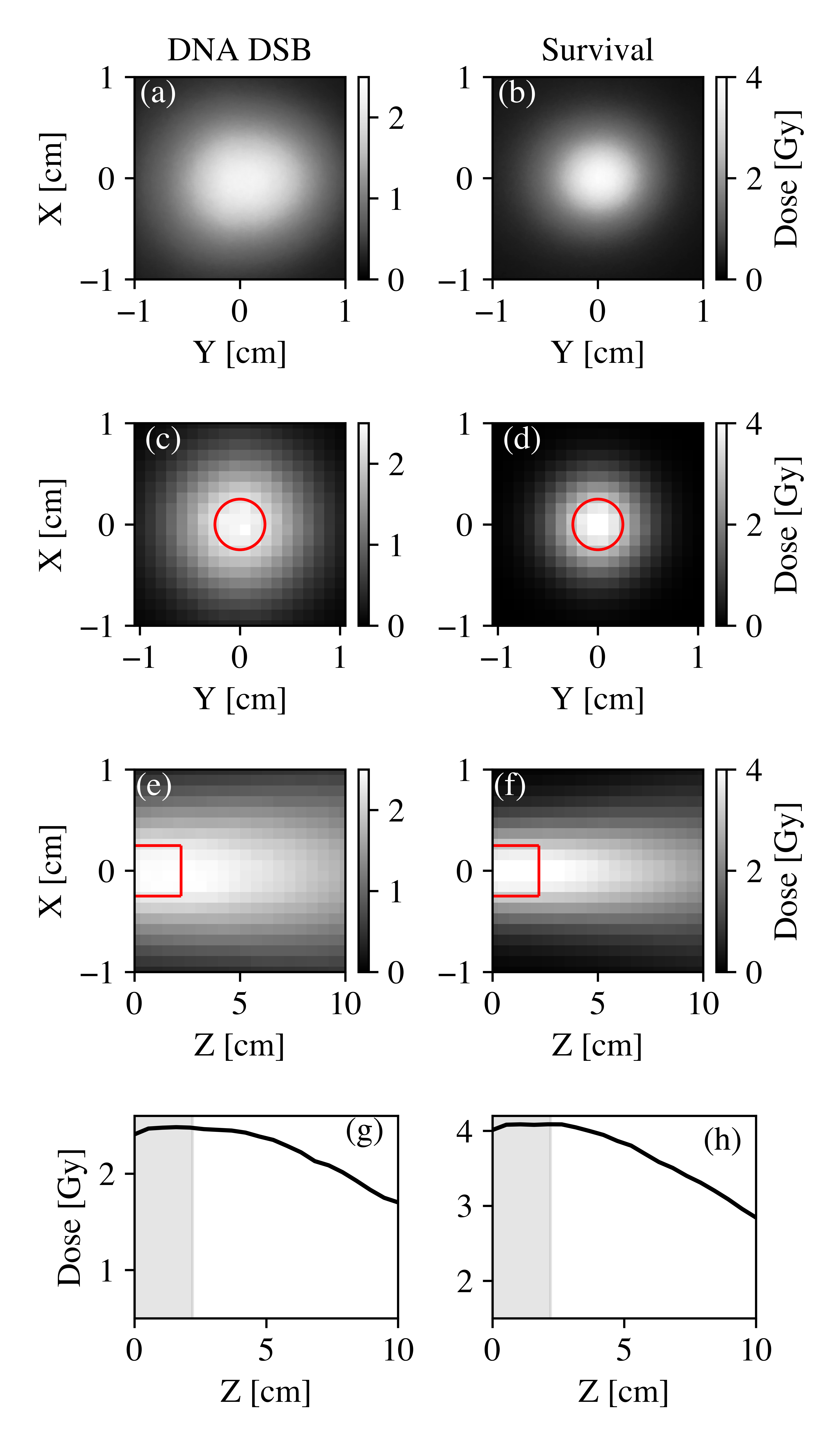}
\caption{\textbf{Experimental and simulated dose profiles:} \textbf{(a)},\textbf{(b)} Examples of the measured transverse profiles of the delivered dose for the foci formation and clonogenic assays studies, respectively. \textbf{(c)},\textbf{(d)} Corresponding simulated dose profiles obtained using the Monte-Carlo scattering code TOPAS and assuming the average electron spectra shown in Fig. \ref{fig3}. Red circles indicate the area occupied by the cell cultures. \textbf{(e)},\textbf{(f)} Simulated dose-depth profiles for each configuration. The solid rectangles indicate the regions occupied by the cells. \textbf{(g)},\textbf{(h)} Lineout along the central axis of the simulated dose-depth profiles shown in frames (e) and (f), respectively. The grey shaded regions indicate the regions occupied by the cell cultures.}
\label{fig4}
\end{figure}

Figs. \ref{fig4}(a) and \ref{fig4}(b) show the typical spatial distribution of the dose delivered at the cell plane for both the foci formation and the clonogenic studies, respectively.
The dose profile is well approximated by a Gaussian distribution with a standard deviation of \unit[0.6]{cm} and \unit[0.4]{cm} for the two cases. In the cell region, the typical coefficient of variation of the dose distribution is 8.4\% and 14.1\%, respectively. 

The observed dose was confirmed by matching Monte-Carlo simulations using the TOPAS (Geant4) code \cite{perl2012topas, agostinelli2003geant4, allison2016recent} (compare Figs. \ref{fig4}(a-b) with \ref{fig4}(c-d)). In the simulations, the electron beam was input with a custom energy spectrum replicating the measured electron spectra in Fig. \ref{fig3}.
The source size was taken as \unit[10]{\textmu m}, whereas the beam divergence was set to be 2.3 and \unit[3.6]{mrad} for the foci formation and survival studies, respectively, in order to match the values measured during the experiment.
In the simulations, the cell sample was replaced with a 2$\times$2$\times$\unit[10]{cm} water phantom, split into \unit[50]{\textmu m$^3$} voxels for scoring dose in the transverse and longitudinal direction. 
The simulations were ran with $10^9$ primaries, with the results then scaled to match the  electron beam charge measured during the experiment. 
The simulations also indicate a rather uniform dose-depth profile (Figs. \ref{fig4}(e-h)) with a variation of only 1.2\% over the region occupied by the cells, consistent with the weak dependency of linear energy transfer on particle energy for ultra-relativistic electrons. 
All these uncertainties in the dose have been taken into account in the analysis of the data. While the electron spectra presented some level of shot-to-shot variability due to intrinsic fluctuations in the laser and gas parameters, the results presented in this work are independent of it, thanks to the on-shot measurements of the dose properties for each single-pulse irradiation. 
The area irradiated and the longitudinal and transverse uniformity of the dose delivered are sufficient also for future in-vivo studies \cite{kroll_2022}. Moreover, the shot-to-shot fluctuations observed in this work, while not affecting the results reported here, can in principle be reduced down to the percent level in a similar wakefield accelerator, as recently demonstrated experimentally \cite{Maier}.

\newpage

\section*{Cell cultures and analysis}
Human skin fibroblast cells (AG01522 D) were obtained from Coriell Institute for Medical research and maintained in $\alpha$-modified essential medium (MEM), supplemented with 20\% Fetal Bovine Serum (FBS) and 1\% penicillin-streptomycin. Early passage cells (2-4) were used here after procurement from the Coriell institute. Patient-derived glioblastoma stem-like cells (E2 cells) that expressed the most common stem cells biomarkers such as Nestin and Sox-2 were also irradiated. The cells were cultured in Advanced DMEM-F12 medium supplemented with B27, N2, L-Glutamine, heparin, epidermal growth factor and basal fibroblast growth factor. All cells were incubated in 5\% CO$_2$ in air (unless otherwise stated) with 95\% humidity at \unit[37]{$^\circ$C}. 

For monolayer irradiations (foci kinetics), cells were plated in triplicates for each sample \unit[24]{hours} prior to irradiation inside slide flasks, with approximately $1-3\times10^5$ cells transferred and incubated with \unit[5]{ml} cell-specific medium. Prior to irradiation, the medium was removed and the slide flask filled with fresh medium, as to cover the whole surface and prevent dehydration during the irradiation process. After irradiation, the medium was replaced to cover the cells before incubation, until the desired fixation time had passed.

The preparation procedure was different for cell suspensions (cell survival assay). Prior to irradiation, the cells were detached from their culture flask using Accutase. The cell concentration was then measured and used to fill a \unit[500]{\textmu l} Eppendorf tube. 
A typical cell concentration in this case was \unit[$1.5\times10^5$]{cells/ml}, implying that $\approx 7.5\times 10^4$ cells were irradiated for each sample. Six replicates were used for each endpoint.

In order to assess cellular response under hypoxic conditions, the cells were grown on customised stainless steel dishes with a \unit[3]{\textmu m} Mylar membrane base for cell attachment. The dishes were then inserted in an custom-built hypoxia chamber (see \cite{Chaudhary2022} for full details). \unit[95]{\%} N$_2$ and \unit[5]{\%} CO$_2$ gas was passed through the chambers at a flow rate of \unit[$\approx$0.5] {l/min}. Once disconnected from the gas supply, the chambers were then sealed. Previous calibration showed that the oxygen concentration remained below 0.4\% O$_2$ for up to 45 minutes \cite{Chaudhary2022}. 

DNA double-strand break (DSB) repair kinetics were measured by the persistence of radiation-induced foci as a function of time after irradiation. After irradiation, the cells were fixed in 5\% paraformaldehyde at time points of 0.5, 1, 6 and \unit[24]{hr}.
After fixation, the cells were permeabilized, blocked in 10\% goat serum and probed with 53BP1 primary antibody for 1 hour at 37 $^\circ$C. The unbound primary antibody was then rinsed away with triple washing of the samples. Subsequently, the cells were labelled with Alexa Fluor 488 conjugated secondary antibody and finally mounted with glass coverslips using prolong gold antifade reagent containing DAPI and left overnight to dry. For hypoxia samples, we used an antibody specific to hypoxia inducible factor 1-$\alpha$ and used the same immuno-staining steps used for 53BP1.
In each condition, 2 sham irradiations were performed to measure endegenous and background foci formation. The oxic control values were 3.4 $\pm$ 1.0 and 2.0 $\pm $1.3 for AG0 and E2 respectively, with similar values of 2.9 $\pm$ 1.1 and 3.0 $\pm$ 0.9 for hypoxic conditions. The number of 53BP1 foci reported here are radiation-induced and therefore control values had been subtracted from the measured values. For comparison of results between Mylar and slide flask grown cells, a reference comparison was performed at \unit[0.5]{hr} post irradiation. It was found that under oxic conditions on Mylar and slide flasks the radiation-induced foci/cell/Gy were 24.3 $\pm$ 1.4 and 25.2 $\pm$ 1.6 respectively. Therefore, there is no statistically significant effect on DNA DSB damage for cells grown on slide flasks or Mylar, allowing a direct comparison between hypoxic and oxic data-sets despite the different cellular set-ups. 

Cell survival assays were also carried out to quantify the ability of a cell to reproduce and form colonies 12-14 days after irradiation.  
After irradiation, the cells were transferred from suspension to six well-plates along with fresh medium before incubation. 
After 14 days, the cell medium was removed and the cells were stained with crystal violet dye \cite{feoktistova2016crystal}.
The survival fraction (SF) is then quantified as
\begin{equation}
\text{SF} = \frac{\text{Colonies}}{\text{Cells plated} \times \text{PE} }   
\end{equation}
where \emph{PE} is the plating efficiency, which was measured during two sham irradiations as 9.8\% and 11.2\% for AG0 and E2 cells, respectively.

\section*{Foci formation results}
\begin{figure*}[t]
\centering
\includegraphics[width=1\linewidth]{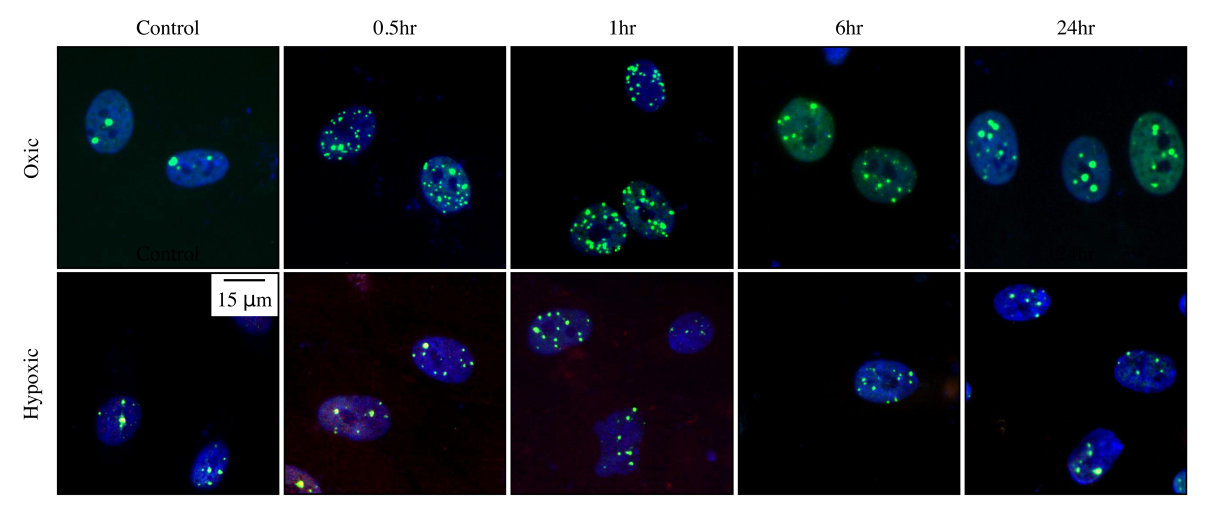}
\caption{\textbf{53BP1 foci distributions as a function of time after irradiation for AG01522 D:} Examples of merged channel images of AG01522 D samples stained with 53BP1 DNA DSB marker (green), DAPI nuclear stain (blue) and HIF-1$\alpha$ hypoxia marker (red). The images are shown as (from left to right) control, 0.5, 1, 6 and \unit[24]{hr} after irradiation and for oxic and hypoxic conditions (Top and bottom, respectively). Split channel images for all irradiation conditions and for both cell lines are available as supplementary material \cite{suppl}.}
\label{fig5}
\end{figure*}

DNA double-strand break repair kinetics were measured using 53BP1 foci induction for control, 0.5, 1, 6 and \unit[24]{hours} after irradiation under both hypoxic and oxic conditions and for both cell types (E2 and AG01522 D). As an example, Fig.\ref{fig5} shows typical images of stained cells, as a function of time after irradiation for AG01522 D cells, with a direct comparison between oxic and hypoxic conditions. The nuclear region of the cell is visible in blue, showing consistent nuclear size and spacing between samples. Exemplary datasets for all irradiation conditions and for both cell lines are available as supplementary material \cite{suppl}. 

\begin{figure}[H]
\centering
\includegraphics[width=\linewidth]{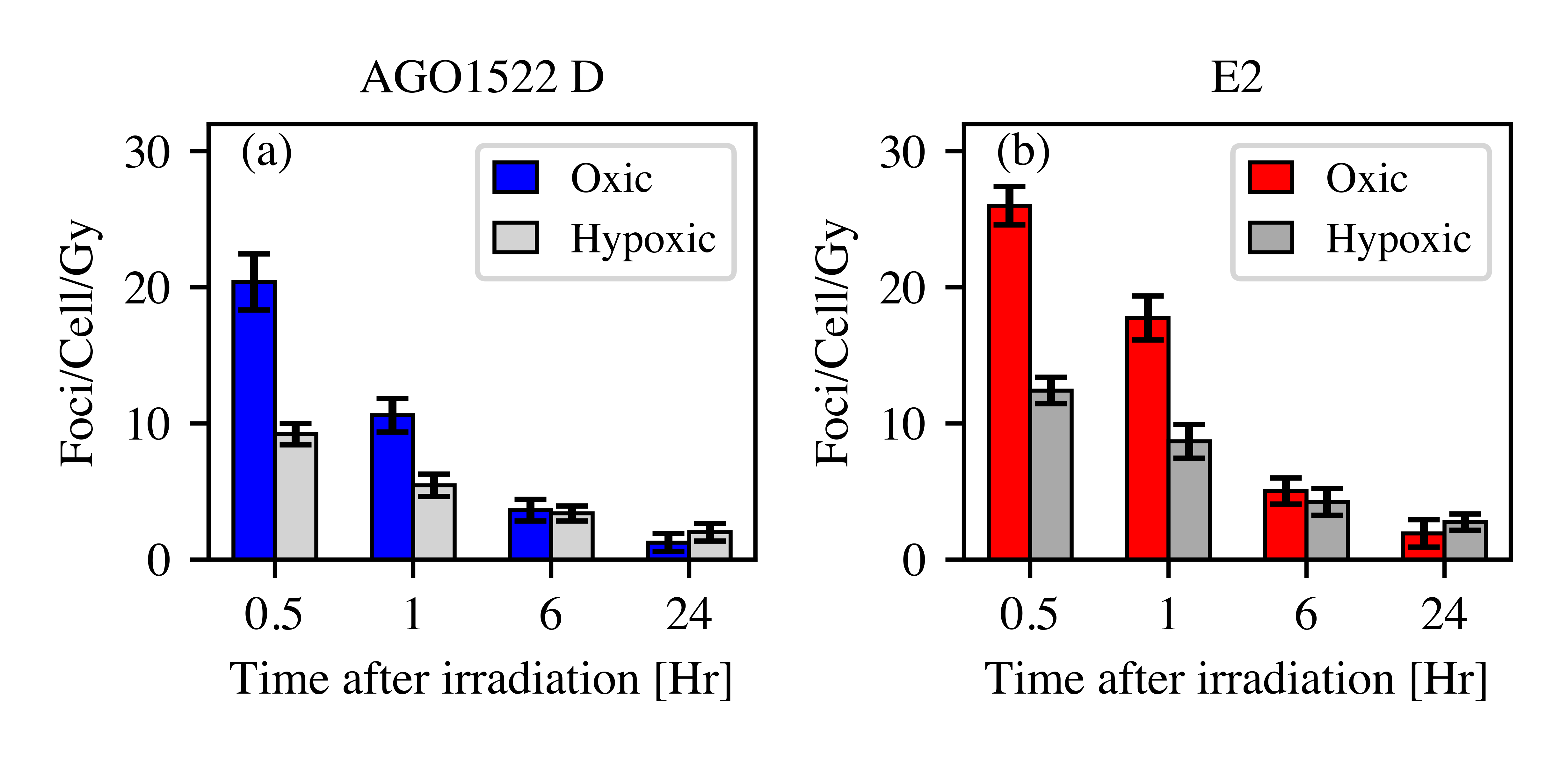}
\caption{\textbf{53BP1 foci measurements for oxic and hypoxic conditions:} Background-corrected 53BP1 foci in AG01522 D \textbf{(a)} and E2 \textbf{(b)} cell samples as a function of time after electron irradiation. For each cell line, 53BP1 foci are measured in hypoxic ($\simeq$\unit[0]{\%} $O_2$) and normoxic  ($\simeq$\unit[20]{\%} $O_2$) conditions. Each distribution is obtained from a population of $>$150 irradiated cells in two individual repeats, with error bars representing the standard deviation.}
\label{fig6}
\end{figure}

Up to one hour after irradiation, an expected reduction of foci in the hypoxic case for both cell lines is observed, a phenomenon linked to the well-known radiosensitising role played by oxygen in biological cells, confirming successful hypoxia induction in our set-up. Successful induction of hypoxia is also visible in the split channel images, available as supplementary material \cite{suppl}, for both the AG0 and E2 cells. 
As expected, the number of 53BP1 foci decreases in time for all conditions. However, the number of foci still persisting 24 hours after irradiation is not significantly different between the oxic and hypoxic conditions, at approximately 2 foci per cell per Gy (Fig.\ref{fig6}). 

\begin{figure}[H]
\centering
\includegraphics[width=\linewidth]{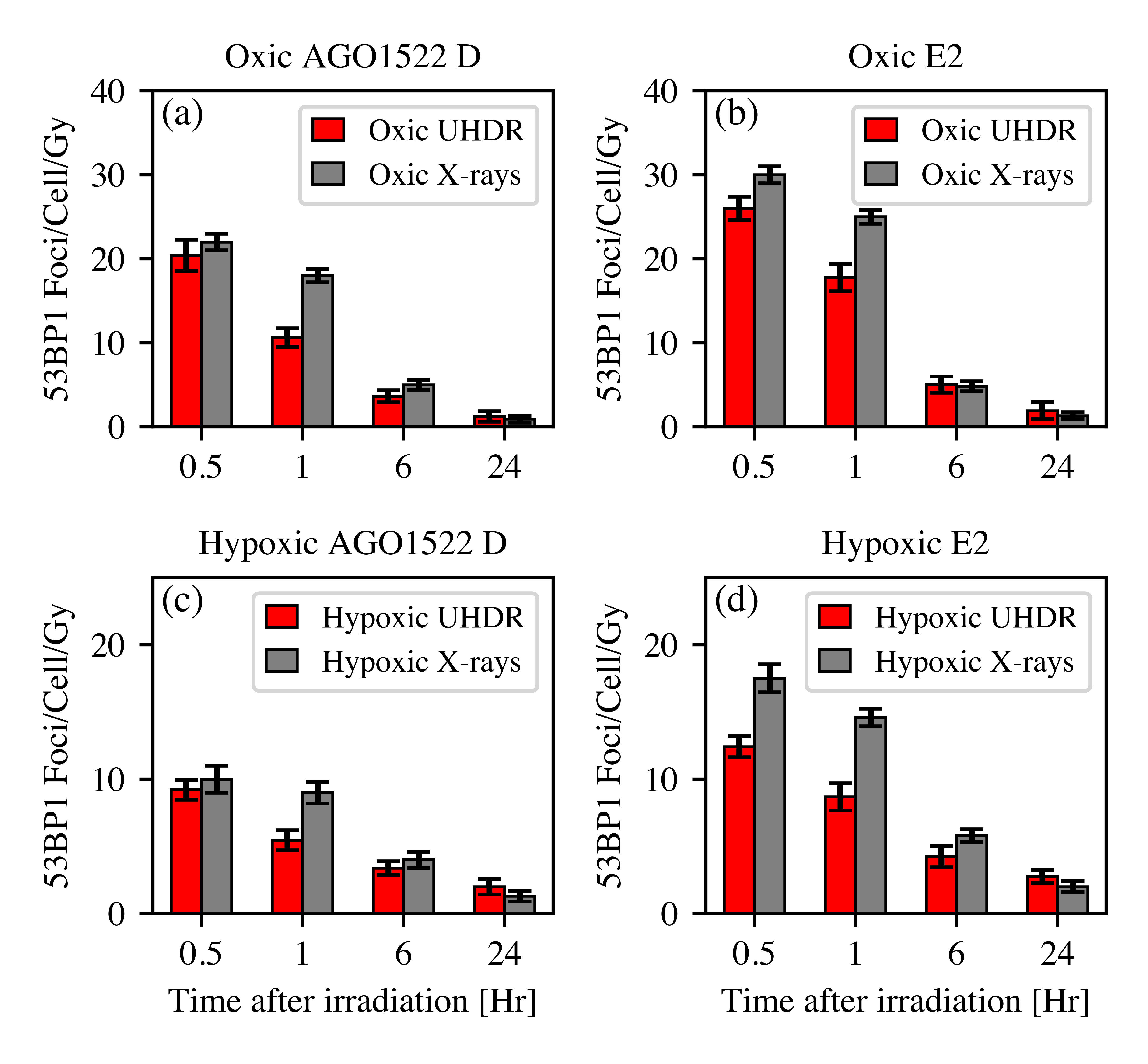}
\caption{\textbf{Comparison of UHDR radiation-induced 53BP1 foci with x-ray data:} In all frames, red and grey bars represent the results for UHDR irradiation and for irradiation using the low dose-rate x-ray source, respectively. \textbf{(a)} Oxic conditions for the AG01522 D cells; \textbf{(b)} oxic conditions for the E2 cells; \textbf{(c)} hypoxic conditions for the AG01522 D cells; and \textbf{(d)} hypoxic conditions for the E2 cells. Each distribution is obtained from a population of $>$150 irradiated cells in two individual repeats, with error bars representing the standard deviation.}
\label{fig7}
\end{figure}

In order to quantify potential differences from lower dose-rate irradiations, reference experiments have been carried out in similar conditions using a \unit[225]{kVp} x-ray source (X-Rad 225, Precision, USA), operating at a dose-rate of \unit[0.49]{Gy/min}. 
Also in this case, both cell types were irradiated in oxic and hypoxic conditions (split channel images available as supplementary material \cite{suppl}). A direct comparison between the two dose-rates for all the different configurations is shown in Fig. \ref{fig7}. In the oxic case (frames a. and b.), both AG0 and E2 show a similar amount of foci induced immediately after irradiation for the two dose-rates. No statistically significant difference is observed also in the average number of foci persisting 24 hours after irradiation, which is of the order of 1-2 foci per cell per Gy in all cases.

In the hypoxic case, there is an expected overall reduction in the number of foci detected promptly after irradiation. However, the average number of foci persisting 24 hours after irradiation does not show any significant difference between the two cell lines and the two dose-rates. 

It must be noted that a comprehensive characterisation of the role of oxygen concentration in cell damage and repair mechanisms following femtosecond-scale irradiation will require dedicated and extensive studies (see also the Discussion section). However, this proof-of-concept assessment gives a preliminary indication that oxygen tension might play a limited role in the response of the cell to femtosecond-scale radiation and demonstrates the full viability of this experimental platform to carry out systematic studies in this area.

\begin{figure}[b!]
\centering
\includegraphics[width=\linewidth]{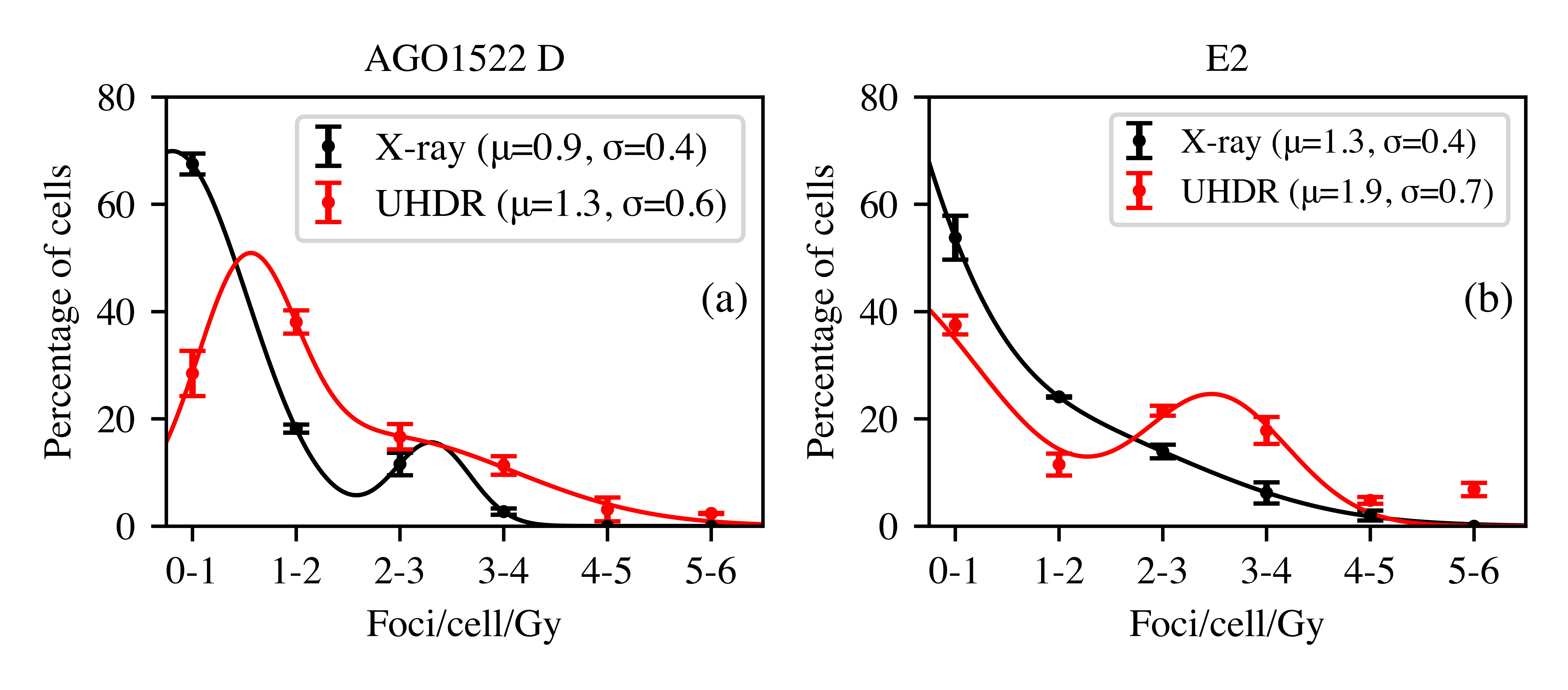} 
\caption{\textbf{Distribution of sub-population of cells as a function of foci persisting 24 hours after irradiation} for \textbf{(a)} AG01522 D and \textbf{(b)} E2, in oxic conditions. Each distribution is obtained from a population of $>$150 irradiated cells in two individual repeats, with error bars representing the standard deviation in the data set. A two-Gaussian fit has been applied to guide the eye. The inset indicates the average number ($\mu$) and the standard deviation in the sample ($\sigma$) in each condition.}
\label{fig8}
\end{figure}

While there is no statistically significant difference in the average number of foci persisting 24 hours after irradiation for the UHDR irradiations, a significant difference is observed in the sub-populations of cells exhibiting multiple foci. This is shown in Fig. \ref{fig8}, which depicts histograms of the number of cells as a function of the number of radiation-induced foci persisting 24 hours after irradiation, for both AG01522 D and E2 cells. 
For each cell-line, the distribution of foci per cell in UHDR conditions is significantly different from that obtained at more conventional dose-rates, with an apparent increase in the number of cells with more than one damage site (71$\pm$4\% for the UHDR compared with 33$\pm$2\% for x-ray irradiation for the AG01522D cells and 63$\pm$2\% versus 46$\pm$3\% for the E2 cells). A two-sample Kolmogorov-Smirnov test on the distributions yields p-values $<0.001$ in both cases, a strong indication that the two distributions are significantly distinct. 

\begin{figure}[b!]
\centering
\includegraphics[width=\linewidth]{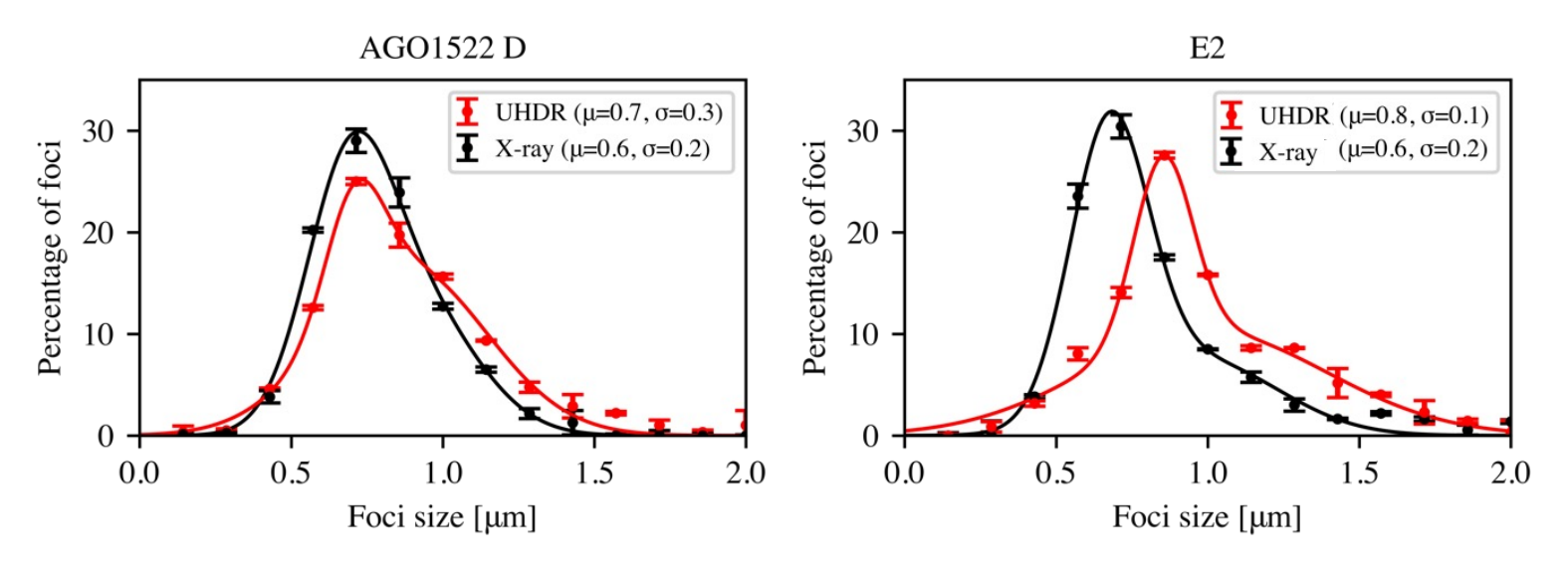}
\caption{\textbf{Distribution of radiation induced 53BP1 foci size 24 hours after irradiation} for \textbf{(a)} AG01522 D and \textbf{(b)} E2 cells in oxic conditions. Each distribution is obtained from $>100$ foci in two repeats, with error bars representing the standard deviation in the data set. A two-Gaussian fit has been applied to guide the eye. The inset indicates the average size (\textmu) and the standard deviation in the sample ($\sigma$) in each condition.}
\label{fig8a}
\end{figure}

The distributions of the size of the damage sites \unit[24]{hr} after irradiation are shown in Fig. \ref{fig8a} for both cell lines under both irradiation conditions. The data is well-fitted by a two-Gaussian distribution in all cases, indicating the presence of a bulk of relatively small-size foci and of a smaller population of foci with a higher size. An increase in the average size of the damage sites is observed for the UHDR irradiation for both AG01522 D and E2 cells, even though it is less pronounced for the AG01522 D cells (see Fig. \ref{fig8a}). In both cases a higher population of damage sites with sizes larger than 1 $\mu$m contributes to the increase in average size. A two-tailed t-test on both cell lines yields a p-value $< 0.001$, indicating a high significance for the increase in average foci size for both cell lines. 

In summary, while the total number of damages is not significantly different in the UHDR irradiations (Fig. \ref{fig7}), the increase in sub-population of cells with a higher number of damages (Fig. \ref{fig8}) and in the size of the damage sites (Fig. \ref{fig8a}) are indicative of more structured and clustered damages. This is a preliminary signature of the induction of more complex - and, thus, less likely to be repaired - damages induced by UHDR irradiation, with an indication of this effect being more pronounced for E2 cells. Direct comparison between oxic and hypoxic conditions indicates that this effect might not be dominated by oxygen depletion induced by the UHDR irradiation (more details in the Discussion section).


\section*{Clonogenic assay results}
\begin{figure}[H]
\centering
\includegraphics[width=\linewidth]{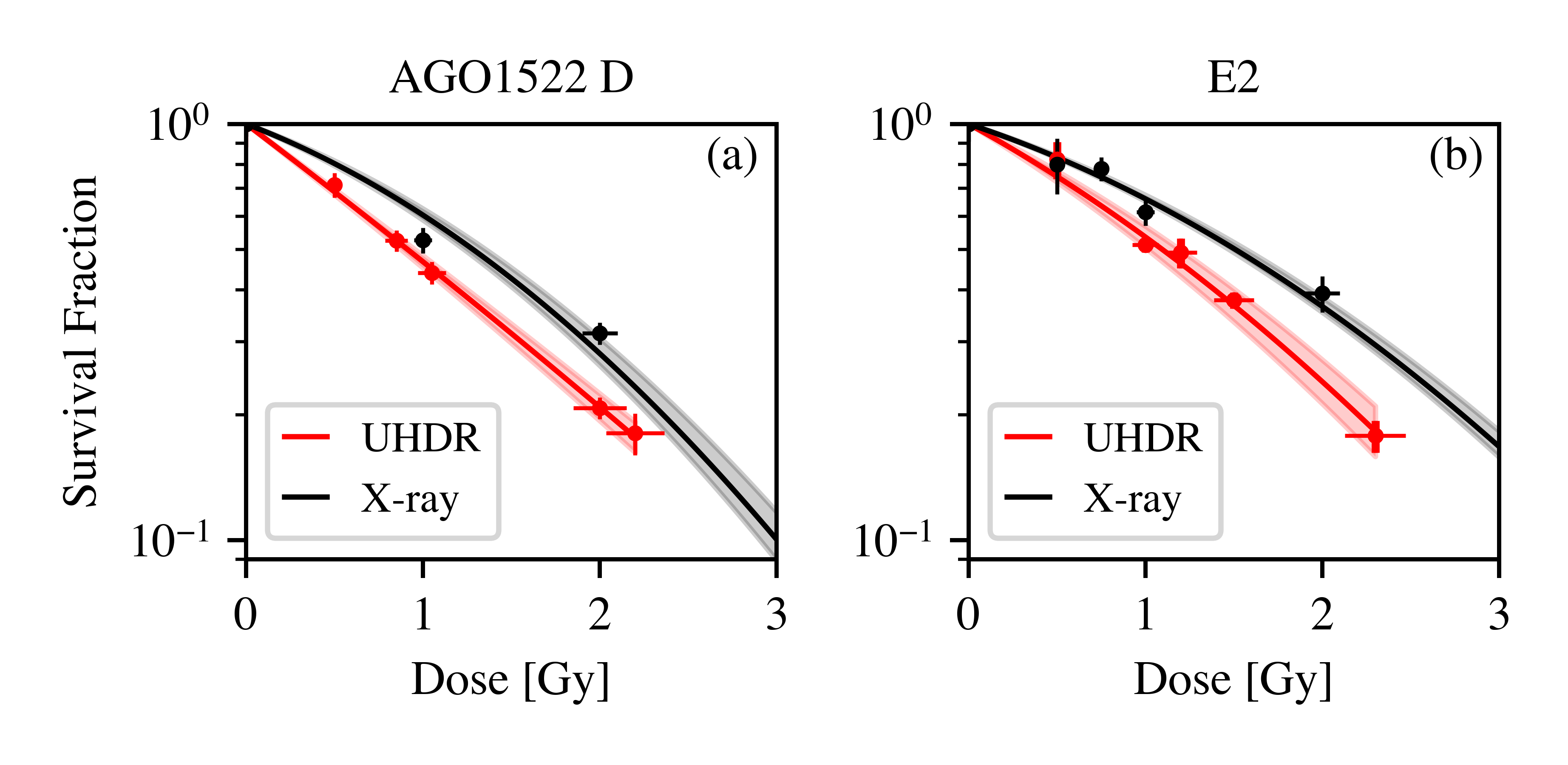}
\caption{\textbf{Cell survival assay results}: Comparison of cell survival assay for \unit[225]{kVp} x-rays (black) and wakefield accelerated electrons (UHDR, red) for \textbf{(a)} AG01522 D and \textbf{(b)} E2 cells. Each data set is fitted using the linear quadratic model (Eq. \ref{eq1}). The error in the survival fraction is estimated as the standard error in the mean from two sets of six repeated measurements. The error in dose is estimated as the convolution of the coefficient of variation across the irradiation region and uncertainties in the dose measurements reported above. X-ray data is available as supplementary material \cite{suppl} up to \unit[4]{Gy} but shown here only up to \unit[3]{Gy}, for easier visual comparison between x-ray and UHDR irradiations.}
\label{fig9}
\end{figure}

Cell survival assays in oxic conditions were performed to complement the DNA DSB repair kinetics discussed in the previous section. 
The survival fraction as a function of delivered dose is shown in Fig. \ref{fig9} for both cell types, with data in red showing the UHDR results and data in black showing reference data obtained with the 225 kVp x-ray source operating at a dose-rate of \unit[0.49]{Gy/min}. The data is shown up to \unit[3]{Gy} for a better visual comparison between x-ray and UHDR irradiation; full data sets for the x-ray irradiation over a larger dose range are available as supplementary material \cite{suppl}.
Each dataset is fitted using the linear-quadratic model \cite{mcmahon2018linear}:
\begin{equation}
S(D) = \exp{-(\alpha D + \beta D^2)},\label{eq1}
\end{equation}

where $S(D)$ represents the survival fraction as a function of delivered dose $D$ and $\alpha$ and $\beta$ are fitting parameters, qualitatively representing the relative weight of prompt damage induced by clustered hits from a single track and accumulation of damage from multiple tracks, respectively \cite{mcmahon2018linear,kogel}. The obtained fitting parameters and their associated uncertainties for each data set are shown in Table \ref{table}.

\begin{table*}[b!]
\centering
\begin{tabular}{|c|c|c|c|c|c|c|}
\hline
 Cell line & Source & $\alpha (Gy^{-1})$ & $\beta (Gy^{-2})$ & $\Bar{D} (Gy)$ & RBE (50\%) & $\chi^2$ \\
 \hline
 \hline
 \multirow{2}{1.5cm}{\hspace{0.35cm}AG0} & x-ray & 0.37 $\pm$ 0.03 & 0.13 $\pm$ 0.01 & 1.46 $\pm$ 0.08 & & 3.3\\
 \cline{2-7}
  & $e^-$ &  0.74 $\pm$ 0.01  &  0.02 $\pm$  0.01 & 1.27 $\pm$ 0.05 & 1.41 $\pm$ 0.08 & 0.5\\ 
  \hline
  \hline 
 \multirow{2}{1.5cm}{\hspace{0.5cm}E2} & x-ray & 0.32 $\pm$ 0.01 & 0.09 $\pm$ 0.01 & 1.77 $\pm$ 0.02 &  & 2.3\\
 \cline{2-7}
  & $e^-$ & 0.54 $\pm$ 0.06 &  0.09 $\pm$ 0.02  & 1.35 $\pm$ 0.04 & 1.40 $\pm$ 0.08 & 1.0\\ 
 \hline
\end{tabular}
\caption{Table of fitting parameters for figure \ref{fig9} together with the mean inactivation dose $\Bar{D}$, and the relative biological effectiveness (RBE) at 50\%. The reduced $\chi^2$ values are provided as a measure of the goodness of fit. }
\label{table}
\end{table*}

The reference data obtained at \unit[0.49]{Gy/min} shows an expected degree of enhanced radioresistance for tumour cells, with comparatively lower values of both $\alpha$ and $\beta$. This results in a higher mean inactivation dose with $\bar{D}=(1.77\pm0.02)$ Gy for E2 and $\bar{D}=(1.46\pm0.08)$ Gy for AG01522 D. However, data obtained at UHDR show a significantly different behaviour. The first apparent difference is a lower survival fraction for both cell lines across the dose range reported in this study, resulting in a relative biological effectiveness (RBE) of $1.40\pm0.08$ at 50\% survival for both cell lines. Associated with this, we observe a significantly lower mean inactivation dose for both cell lines, when compared to irradiations at \unit[0.49]{Gy/min} (i.e., $\bar{D}=(1.35\pm0.04)$ Gy versus $\bar{D}=(1.77\pm0.02)$ Gy for E2 and $\bar{D}=(1.27\pm0.05)$ Gy versus $\bar{D}=(1.46\pm0.08)$ Gy for AG01522 D). The lower survival fraction observed for the UHDR electron irradiation appears to be related to a significantly higher $\alpha$ parameter, possibly indicating a higher probability of induction of lethal clustered damages. Interestingly, the almost purely linear shape of the UHDR survival curves closely resembles those typically obtained with irradiation with higher linear energy transfer (LET) particles, such as ions or protons (see, e.g., Ref. \cite{doria2012biological}). This is also suggestive of more complex damage from a single track. 

Another difference observed at these dose-rates is the significant reduction in the relative radioresistance of the E2 cells (see also Fig. 1b in the supplementary material \cite{suppl} for a direct comparison between E2 and AG01522D at UHDR). This is demonstrated by a similar mean inactivation dose for the two cell lines at UHDRs with a relative difference of $(\bar{D}_{E2}-\bar{D}_{AG0})/\bar{D}_{E2}\simeq (6\pm3)\%$ for the UHDR irradiations, to be compared with a relative difference of $\simeq(18\pm3)\%$ for the low dose-rate irradiations. As an example of this reduced radioresistance of E2 cells, AG01522 D and E2 cell present a similar suvival fraction at \unit[2.2]{Gy}, the highest dose accessed in this work (i.e., ($16\pm4$)\% for AG01522 D and ($18\pm2$)\% for E2). 


\section*{Discussion and Conclusions}
The results presented here give the first experimental demonstration of the full capability of laser-wakefield accelerators to access a novel regime of radiobiology, whereby single-shot Gy-scale doses can be delivered, with a high degree of spatial uniformity, over timescales matching those of the physical response to biological cells to radiation. Full on-shot characterisation of the dose delivered allows disentangling of the results from any potential shot-to-shot variations in the electron beam characteristics. 
The irradiation area and dose-depth profile achievable by the accelerator are shown to be sufficient to proceed to follow-up in-vivo studies (similar to Ref. \cite{kroll_2022}). 
Full characterisation of the accelerator indicates that tuneable doses up to \unit[2.2]{Gy} can be delivered over timescales of the order of \unit[150]{fs}, resulting in unprecedented dose-rates up to \unit[$10^{13}$]{Gy/s}. Numerical modelling indicates that further optimisation of the laser and plasma parameters can allow reaching doses of up to \unit[4]{Gy} per pulse with a 100-TW class laser, with higher doses in principle achievable at PW-scale laser facilities.

The viability of the radiation source to carry out radiobiological experiments has been tested with pilot irradiation of patient-derived glioblastoma stem-like cells (E2 cell line) and human skin fibroblast cells (AG01522 D cell line), typical models to mimic tumour and healthy tissue in the laboratory. The results have then been compared with irradiation using a \unit[225]{kVp} x-ray source with a dose rate of \unit[0.49]{Gy/min}, a typical standard in comparative radiobiology with a defined RBE = 1.0 \cite{Murshed}. Preliminary indications of differential biological response to femtosecond-scale radiation have been observed, providing hypoteses-generating evidence for further studies in this regime, in conditions that are fundamentally different from FLASH radiotherapy.

Clonogenic assays indicate a statistically significant decrease in survival rate for both cell types, when compared to the same range of doses delivered at \unit[0.49]{Gy/min}. This is exemplified by a net decrease in mean inactivation dose and an increase in the relative biological effectiveness for both cell types (RBE = $1.40\pm0.08$ at 50\% survival). The data also evidence an apparent relative reduction of radioresistance for the E2 tumour cells that, for doses of the order of \unit[2]{Gy}, present similar survival rates when compared to the AG01522 D cells (survival fractions of $18\pm2$\% and $16\pm4$\% for E2 and AG01522 D, respectively, see also Fig. 1b in the supplementary material \cite{suppl}). The high RBE is accompanied by an increase in the $\alpha/\beta$ ratio, a first indication of the induction of more complex damage in the cell.

The observed increase in radiobiological effectiveness does not appear to be correlated with the average number of foci persisting up to 24 hours after irradiation, which are not significantly different from those observed during low dose-rate irradiations. However, analysis of the sub-populations of cells does indicate a clear increase in the percentage of cells with multiple radiation-induced foci persisting 24 hours after irradiation. An associated and statistically significant increase in the average size of the damage sites is also observed. 

Recent experiments on the same cell cultures using 20-ps long electron beams did not provide clear evidence of the same effects \cite{mcanespie2023ijrobp}, similarly to independent results using ps-long x-ray beams \cite{hill2002increased}. While it is worth noting that the experiments reported in Ref. \cite{mcanespie2023ijrobp} suffered of significant uncertainties in dose due to the structured dose-depth profile of MeV-scale electrons \cite{McAnespie_pre}, their broad agreement with the results reported in Ref. \cite{hill2002increased} provides additional indications that such differential response might be distinctive of femtosecond-scale irradiation. 

\begin{figure}[H]
\centering
\includegraphics[width=\linewidth]{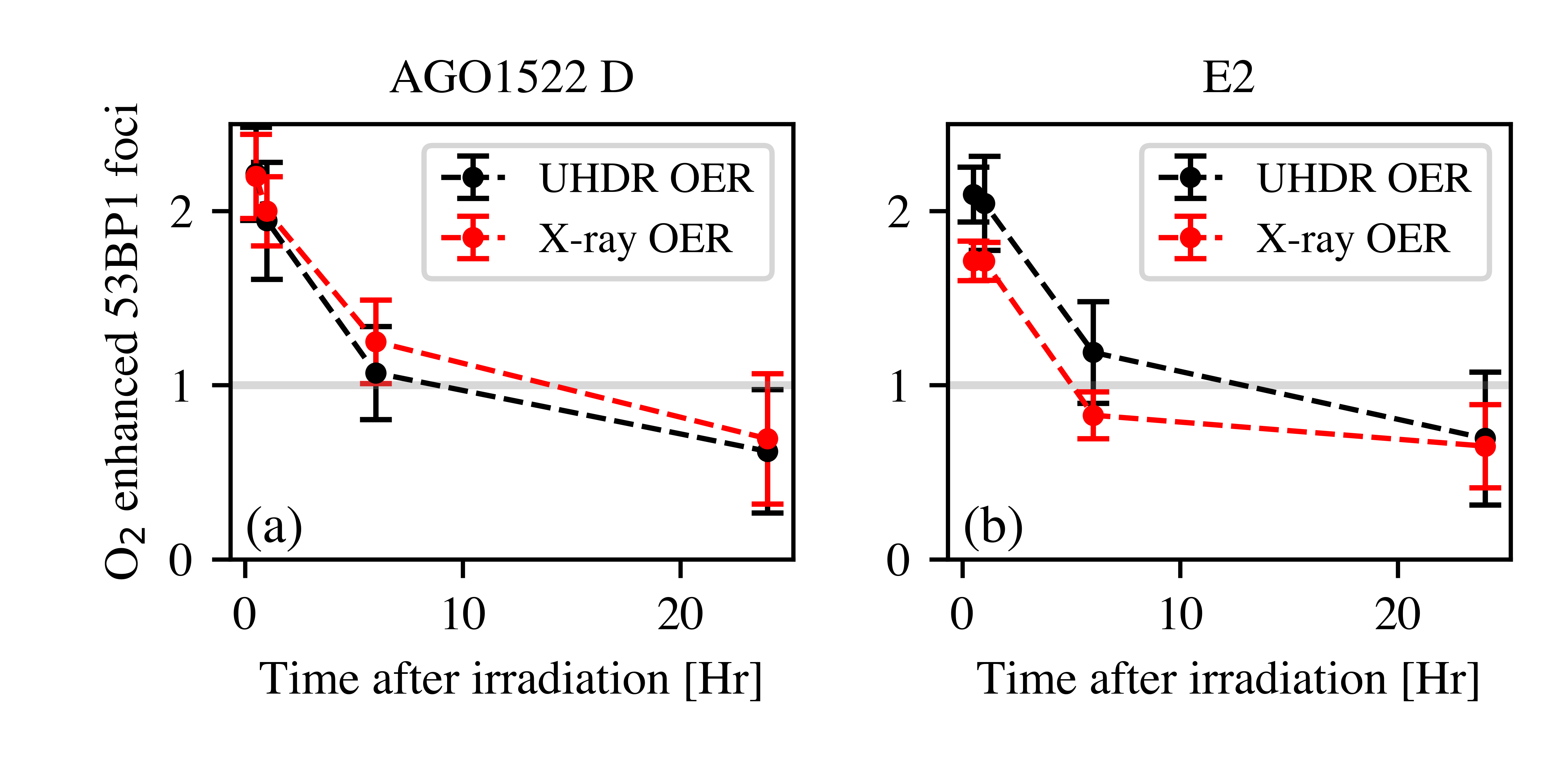}
\caption{\textbf{Measured oxygen enhancement foci induction for (a) E2 glioblastoma stem-like cells and (b) AG01522 D human skin fibrobalsts cells:} In both graphs, red and black lines refer to the UHDR and reference x-ray data, respectively.}
\label{fig10}
\end{figure}

A direct comparison between oxic and hypoxic foci formation results does not evidence a clear link between this reduction in tumour radioresistance and the oxygen levels in the cells. This is illustrated in Fig. \ref{fig10}, which shows broad agreement between the oxygen enhancement ratios (OER) measured in both dose-rate regimes and for both cell lines.
The figure shows that, 30 minutes after irradiation, the number of foci is consistent with the OER quoted in literature for low LET radiation sources \cite{kogel}. While oxygen depletion is one of the possible causes for the observed increase in radioresistance for normal cells irradiated under FLASH conditions \cite{wilson2020ultra,jansen2021does}, our results do not show clear evidence that oxygen depletion might play a significant role in the differential effects of femtosecond-scale irradiations reported here. This can be qualitatively understood if we consider that, both these and previously reported picosecond irradiations \cite{mcanespie2023ijrobp} deliver their dose in a timescale much shorter than the $\approx$ \unit[5]{ms} timescale of oxygen fixation of DNA \cite{watts1978fast}, suggesting that oxygen depletion kinetics are likely to be similar in these two cases, unless other currently unknown physico-chemical effects are involved at these extreme timescales. 

It is not clear what might be the primary cause of the observed increase in radiobiological effectiveness and reduction in radioresistance for tumour cells observed at these dose-rates, and this is the subject of an ongoing and extensive program of work that is beyond the scope of the hypotheses-generating evidence reported here. However, two main, and possibly interlinked, hypotheses can be put forward at this stage.

A first possibility is related to the high volumetric density of a high-charge femtosecond-scale electron beam. At the cell plane, the electron beams used in our experiments had a typical diameter of the order of \unit[1]{cm}, a duration of the order of \unit[150]{fs}, and contained a total charge in the range of 1-\unit[2]{nC}, resulting in a volumetric number density at the cell plane of the order of $10^{12}-10^{13}$ electrons/cm$^3$. This value is approximately two orders of magnitude higher than in the case of picosecond irradiations reported in Refs. \cite{McAnespie_pre,mcanespie2023ijrobp}, and several orders of magnitude higher than in the case of conventional or FLASH radiotherapy \cite{schuler2022ultra}.
The much higher density of ionising tracks allows for a significantly higher probability of non-linear interactions between radical clouds emerging from separate tracks and greatly enhances the probability of multiple neighbouring damages in the DNA. For example, the typical diffusion length for solvated electrons is between 500 nm and 1 micron, whereas the typical diffusion length for OH radicals is between 100 and 400 nm \cite{Roots1975}. Within these distances, the average number of primary electrons is of the order of 1 - 10 for the ultra-high dose-rate irradiation reported here, indicating a high probability of overlap of radical clouds. For comparison, the corresponding average number of electrons for the picosecond-scale irradiation reported in Refs. \cite{McAnespie_pre,mcanespie2023ijrobp} would be between 10$^{-2}$ and 10$^{-1}$. It is thus possible that, while the number of damages induced in the cell DNA is similar to that at lower dose-rates, the complexity of prompt damage might be higher, resulting in a lower probability of efficient repair. This interpretation is consistent with the observed increase in the size of the damage sites, with the increase of the sub-population of cells showing multiple damages, and with the increase of the $\alpha/\beta$ ratio in the survival curves, which is qualitatively associated with cell death by clustered and prompt damages \cite{kogel}.   

Another possible or concurrent explanation is that femtosecond-scale irradiations imply that no radiation is present during radical diffusion and the first onset of chemical reactions. For example, it is established that electrons solvate in water on timescales of the order of 200 - \unit[500]{fs} \cite{Low2022} and that the excitation of solvated electrons might alter their reactivity. An irradiation time of \unit[150]{fs} would thus imply that the radiation does not deposit energy during the electron solvation and diffusion, or any other early radiation-induced physico-chemical phenomena. This can also lead to a highly heterogeneous regime where a very large number of radical products are generated promptly, rather than having a comparatively slow production during more prolonged irradiations. This irradiation modality thus provides a step-change capability compared to ps-scale irradiation previously reported \cite{mcanespie2023ijrobp, McAnespie_pre}, enabling, for the first time, access to the physical stage of cellular response to radiation and delivering a density of ionising tracks sufficient to trigger interactions between radical clouds. 

At present, it is not known what, if any, impact this may have on the subsequent complex radical organic chemistry which will occur within the cell and its subsequent impact on cell survival. The results presented here provide a novel platform and hypotheses-generating evidence for systematic studies in the area. As a result of this knowledge gap, it is currently difficult to generate robust physico-chemical models of these effects at femtosecond timescales and ultra-high dose rates. Monte Carlo radiation transport and chemistry codes have been mainly developed for relatively low dose-rate regimes in liquid water, where tracks are widely dispersed. While some preliminary work has been done to develop inter-track models (see, e.g., Ref. \cite{Thompson}), these are usually based on assumptions only validated at lower dose-rates, where particle densities are much lower than those seen in these UHDR exposures. In addition, typical numerical codes (such as Geant4-DNA and TOPAS-nBio \cite{agostinelli2003geant4,perl2012topas}) are currently under-constrained and show large variations in predicted biological outcome as a function of, e.g., direct damage threshold, chemical stage length, and hydroxyl damage probability \cite{hongyu}. Extensive studies of cellular response to femtosecond-scale radiation will allow for a methodic benchmarking of these codes, in regimes where secondary effects of protracted irradiation are excluded.

The work presented here pave the way for extensive investigations in this area, including, e.g., studies of different cell-lines and transition to in-vivo, quantifying different bio-markers to assess the relative role of non-homologous end-joining and homologous recombination DNA repair pathways, analysing cell cycle progression, cell survival and death pathways, and assessing possible effects on DNA replication. For a definitive assessment of the role of oxygen tension, our proof-of-concept results demonstrate that these studies can be performed with variable levels of environmental oxygen, including normoxia, physoxia, and different degrees of hypoxia using hypoxic chambers developed by our collaboration \cite{Chaudhary2022}.

\section*{Authors contribution}
GS devised and led the experimental program, with contributions from CAM, PC, SJM, and KMP. The wakefield accelerator was setup and optimised by CAM, MJVS, GS, NB, LC, NC, KF, DJ, BK, AML, JM, and PPR. Cell treatment and post-irradiation analysis was preformed by CAM, PC, SWB, and SN. The data were mainly analyzed by CAM, PC, and GS with contributions from MJVS, SJM, and KMP. JS carried out the numerical modelling of the accelerator. GS, CAM, PC, SJM, SWB, and KMP wrote the manuscript, with input from all the authors.

\section*{Acknowledgements} GS wishes to acknowledge support from the EPSRC (grant numbers EP/V049186/1 and EP/V044397/1). KMP wishes to acknowledge support from Brainwaves NI. SJM is supported by a UKRI Future Leaders Fellowship, grant number MR/T021721/1. M.J.V.S. acknowledges support from the Royal Society (URF-R1221874). The authors are grateful to the CLF staff for their support during the experiment. Patient-derived glioblastoma stem-like cells were kindly provided by Prof. Anthony Chalmers (University of Glasgow, UK) and Prof. Colin Watts (University of Birmingham,UK).


\bibliography{Refs}

\begin{thebibliography}{47}%
\makeatletter
\providecommand \@ifxundefined [1]{%
 \@ifx{#1\undefined}
}%
\providecommand \@ifnum [1]{%
 \ifnum #1\expandafter \@firstoftwo
 \else \expandafter \@secondoftwo
 \fi
}%
\providecommand \@ifx [1]{%
 \ifx #1\expandafter \@firstoftwo
 \else \expandafter \@secondoftwo
 \fi
}%
\providecommand \natexlab [1]{#1}%
\providecommand \enquote  [1]{``#1''}%
\providecommand \bibnamefont  [1]{#1}%
\providecommand \bibfnamefont [1]{#1}%
\providecommand \citenamefont [1]{#1}%
\providecommand \href@noop [0]{\@secondoftwo}%
\providecommand \href [0]{\begingroup \@sanitize@url \@href}%
\providecommand \@href[1]{\@@startlink{#1}\@@href}%
\providecommand \@@href[1]{\endgroup#1\@@endlink}%
\providecommand \@sanitize@url [0]{\catcode `\\12\catcode `\$12\catcode `\&12\catcode `\#12\catcode `\^12\catcode `\_12\catcode `\%12\relax}%
\providecommand \@@startlink[1]{}%
\providecommand \@@endlink[0]{}%
\providecommand \url  [0]{\begingroup\@sanitize@url \@url }%
\providecommand \@url [1]{\endgroup\@href {#1}{\urlprefix }}%
\providecommand \urlprefix  [0]{URL }%
\providecommand \Eprint [0]{\href }%
\providecommand \doibase [0]{https://doi.org/}%
\providecommand \selectlanguage [0]{\@gobble}%
\providecommand \bibinfo  [0]{\@secondoftwo}%
\providecommand \bibfield  [0]{\@secondoftwo}%
\providecommand \translation [1]{[#1]}%
\providecommand \BibitemOpen [0]{}%
\providecommand \bibitemStop [0]{}%
\providecommand \bibitemNoStop [0]{.\EOS\space}%
\providecommand \EOS [0]{\spacefactor3000\relax}%
\providecommand \BibitemShut  [1]{\csname bibitem#1\endcsname}%
\let\auto@bib@innerbib\@empty
\bibitem [{\citenamefont {Ahmad}\ \emph {et~al.}(2012)\citenamefont {Ahmad}, \citenamefont {Duke}, \citenamefont {Jena}, \citenamefont {Williams},\ and\ \citenamefont {Burnet}}]{ahmad2012advances}%
  \BibitemOpen
  \bibfield  {author} {\bibinfo {author} {\bibfnamefont {S.~S.}\ \bibnamefont {Ahmad}}, \bibinfo {author} {\bibfnamefont {S.}~\bibnamefont {Duke}}, \bibinfo {author} {\bibfnamefont {R.}~\bibnamefont {Jena}}, \bibinfo {author} {\bibfnamefont {M.~V.}\ \bibnamefont {Williams}},\ and\ \bibinfo {author} {\bibfnamefont {N.~G.}\ \bibnamefont {Burnet}},\ }\bibfield  {title} {\bibinfo {title} {Advances in radiotherapy},\ }\href@noop {} {\bibfield  {journal} {\bibinfo  {journal} {Bmj}\ }\textbf {\bibinfo {volume} {345}} (\bibinfo {year} {2012})}\BibitemShut {NoStop}%
\bibitem [{Ame()}]{American}%
  \BibitemOpen
  \href@noop {} {\bibinfo {title} {https://www.cancer.org/treatment/treatments-and-side-effects/physical-side-effects/second-cancers-in-adults.html"}}\BibitemShut {NoStop}%
\bibitem [{\citenamefont {Salvo}\ \emph {et~al.}(2010)\citenamefont {Salvo}, \citenamefont {Barnes}, \citenamefont {Van~Draanen}, \citenamefont {Stacey}, \citenamefont {Mitera}, \citenamefont {Breen}, \citenamefont {Giotis}, \citenamefont {Czarnota}, \citenamefont {Pang},\ and\ \citenamefont {De~Angelis}}]{salvo2010prophylaxis}%
  \BibitemOpen
  \bibfield  {author} {\bibinfo {author} {\bibfnamefont {N.}~\bibnamefont {Salvo}}, \bibinfo {author} {\bibfnamefont {E.}~\bibnamefont {Barnes}}, \bibinfo {author} {\bibfnamefont {J.}~\bibnamefont {Van~Draanen}}, \bibinfo {author} {\bibfnamefont {E.}~\bibnamefont {Stacey}}, \bibinfo {author} {\bibfnamefont {G.}~\bibnamefont {Mitera}}, \bibinfo {author} {\bibfnamefont {D.}~\bibnamefont {Breen}}, \bibinfo {author} {\bibfnamefont {A.}~\bibnamefont {Giotis}}, \bibinfo {author} {\bibfnamefont {G.}~\bibnamefont {Czarnota}}, \bibinfo {author} {\bibfnamefont {J.}~\bibnamefont {Pang}},\ and\ \bibinfo {author} {\bibfnamefont {C.}~\bibnamefont {De~Angelis}},\ }\bibfield  {title} {\bibinfo {title} {Prophylaxis and management of acute radiation-induced skin reactions: a systematic review of the literature},\ }\href@noop {} {\bibfield  {journal} {\bibinfo  {journal} {Current Oncology}\ }\textbf {\bibinfo {volume} {17}},\ \bibinfo {pages} {94} (\bibinfo {year} {2010})}\BibitemShut {NoStop}%
\bibitem [{\citenamefont {Adams}\ and\ \citenamefont {Jameson}(1980)}]{adams1980time}%
  \BibitemOpen
  \bibfield  {author} {\bibinfo {author} {\bibfnamefont {G.}~\bibnamefont {Adams}}\ and\ \bibinfo {author} {\bibfnamefont {D.}~\bibnamefont {Jameson}},\ }\bibfield  {title} {\bibinfo {title} {Time effects in molecular radiation biology},\ }\href@noop {} {\bibfield  {journal} {\bibinfo  {journal} {Radiation and Environmental Biophysics}\ }\textbf {\bibinfo {volume} {17}},\ \bibinfo {pages} {95} (\bibinfo {year} {1980})}\BibitemShut {NoStop}%
\bibitem [{\citenamefont {Loh}\ \emph {et~al.}(2020)\citenamefont {Loh}, \citenamefont {Doumy}, \citenamefont {Arnold}, \citenamefont {Kjellsson}, \citenamefont {Southworth}, \citenamefont {Al~Haddad}, \citenamefont {Kumagai}, \citenamefont {Tu}, \citenamefont {Ho}, \citenamefont {March} \emph {et~al.}}]{loh2020observation}%
  \BibitemOpen
  \bibfield  {author} {\bibinfo {author} {\bibfnamefont {Z.-H.}\ \bibnamefont {Loh}}, \bibinfo {author} {\bibfnamefont {G.}~\bibnamefont {Doumy}}, \bibinfo {author} {\bibfnamefont {C.}~\bibnamefont {Arnold}}, \bibinfo {author} {\bibfnamefont {L.}~\bibnamefont {Kjellsson}}, \bibinfo {author} {\bibfnamefont {S.}~\bibnamefont {Southworth}}, \bibinfo {author} {\bibfnamefont {A.}~\bibnamefont {Al~Haddad}}, \bibinfo {author} {\bibfnamefont {Y.}~\bibnamefont {Kumagai}}, \bibinfo {author} {\bibfnamefont {M.-F.}\ \bibnamefont {Tu}}, \bibinfo {author} {\bibfnamefont {P.}~\bibnamefont {Ho}}, \bibinfo {author} {\bibfnamefont {A.}~\bibnamefont {March}}, \emph {et~al.},\ }\bibfield  {title} {\bibinfo {title} {Observation of the fastest chemical processes in the radiolysis of water},\ }\href@noop {} {\bibfield  {journal} {\bibinfo  {journal} {Science}\ }\textbf {\bibinfo {volume} {367}},\ \bibinfo {pages} {179} (\bibinfo {year} {2020})}\BibitemShut {NoStop}%
\bibitem [{\citenamefont {Low}\ \emph {et~al.}(2022)\citenamefont {Low}, \citenamefont {Chu}, \citenamefont {Nie}, \citenamefont {Yusof}, \citenamefont {Prezhdo},\ and\ \citenamefont {Loh}}]{Low2022}%
  \BibitemOpen
  \bibfield  {author} {\bibinfo {author} {\bibfnamefont {P.}~\bibnamefont {Low}}, \bibinfo {author} {\bibfnamefont {W.}~\bibnamefont {Chu}}, \bibinfo {author} {\bibfnamefont {Z.}~\bibnamefont {Nie}}, \bibinfo {author} {\bibfnamefont {M.}~\bibnamefont {Yusof}}, \bibinfo {author} {\bibfnamefont {O.}~\bibnamefont {Prezhdo}},\ and\ \bibinfo {author} {\bibfnamefont {Z.}~\bibnamefont {Loh}},\ }\bibfield  {title} {\bibinfo {title} {Observation of a transient intermediate in the ultrafast relaxation dynamics of the excess electron in strong-field-ionized liquid water},\ }\href@noop {} {\bibfield  {journal} {\bibinfo  {journal} {Nature Communications}\ }\textbf {\bibinfo {volume} {13}},\ \bibinfo {pages} {7300} (\bibinfo {year} {2022})}\BibitemShut {NoStop}%
\bibitem [{\citenamefont {Alizadeh}\ and\ \citenamefont {Sanche}(2012)}]{alizadeh2012precursors}%
  \BibitemOpen
  \bibfield  {author} {\bibinfo {author} {\bibfnamefont {E.}~\bibnamefont {Alizadeh}}\ and\ \bibinfo {author} {\bibfnamefont {L.}~\bibnamefont {Sanche}},\ }\bibfield  {title} {\bibinfo {title} {Precursors of solvated electrons in radiobiological physics and chemistry},\ }\href@noop {} {\bibfield  {journal} {\bibinfo  {journal} {Chemical reviews}\ }\textbf {\bibinfo {volume} {112}},\ \bibinfo {pages} {5578} (\bibinfo {year} {2012})}\BibitemShut {NoStop}%
\bibitem [{\citenamefont {Harrington}(2019)}]{harrington2019ultrahigh}%
  \BibitemOpen
  \bibfield  {author} {\bibinfo {author} {\bibfnamefont {K.~J.}\ \bibnamefont {Harrington}},\ }\bibfield  {title} {\bibinfo {title} {Ultrahigh dose-rate radiotherapy: next steps for flash-rt},\ }\href@noop {} {\bibfield  {journal} {\bibinfo  {journal} {Clinical Cancer Research}\ }\textbf {\bibinfo {volume} {25}},\ \bibinfo {pages} {3} (\bibinfo {year} {2019})}\BibitemShut {NoStop}%
\bibitem [{\citenamefont {Sch{\"u}ler}\ \emph {et~al.}(2022)\citenamefont {Sch{\"u}ler}, \citenamefont {Acharya}, \citenamefont {Montay-Gruel}, \citenamefont {Loo~Jr}, \citenamefont {Vozenin},\ and\ \citenamefont {Maxim}}]{schuler2022ultra}%
  \BibitemOpen
  \bibfield  {author} {\bibinfo {author} {\bibfnamefont {E.}~\bibnamefont {Sch{\"u}ler}}, \bibinfo {author} {\bibfnamefont {M.}~\bibnamefont {Acharya}}, \bibinfo {author} {\bibfnamefont {P.}~\bibnamefont {Montay-Gruel}}, \bibinfo {author} {\bibfnamefont {B.~W.}\ \bibnamefont {Loo~Jr}}, \bibinfo {author} {\bibfnamefont {M.-C.}\ \bibnamefont {Vozenin}},\ and\ \bibinfo {author} {\bibfnamefont {P.~G.}\ \bibnamefont {Maxim}},\ }\bibfield  {title} {\bibinfo {title} {Ultra-high dose rate electron beams and the flash effect: From preclinical evidence to a new radiotherapy paradigm},\ }\href@noop {} {\bibfield  {journal} {\bibinfo  {journal} {Medical physics}\ }\textbf {\bibinfo {volume} {49}},\ \bibinfo {pages} {2082} (\bibinfo {year} {2022})}\BibitemShut {NoStop}%
\bibitem [{\citenamefont {Leavitt}\ \emph {et~al.}(2022)\citenamefont {Leavitt} \emph {et~al.}}]{Leavitt}%
  \BibitemOpen
  \bibfield  {author} {\bibinfo {author} {\bibfnamefont {R.~J.}\ \bibnamefont {Leavitt}} \emph {et~al.},\ }\bibfield  {title} {\bibinfo {title} {Hypoxic tumors are sensitive to flash radiotherapy},\ }\href@noop {} {\bibfield  {journal} {\bibinfo  {journal} {bioRxiv}\ ,\ \bibinfo {pages} {2022.11.27.518083}} (\bibinfo {year} {2022})}\BibitemShut {NoStop}%
\bibitem [{\citenamefont {Thompson}\ \emph {et~al.}(2023)\citenamefont {Thompson}, \citenamefont {Prise},\ and\ \citenamefont {McMahon}}]{Thompson}%
  \BibitemOpen
  \bibfield  {author} {\bibinfo {author} {\bibfnamefont {S.}~\bibnamefont {Thompson}}, \bibinfo {author} {\bibfnamefont {K.}~\bibnamefont {Prise}},\ and\ \bibinfo {author} {\bibfnamefont {S.}~\bibnamefont {McMahon}},\ }\bibfield  {title} {\bibinfo {title} {Investigating the potential contribution of inter-track interactions within ultra-high dose-rate proton therapy},\ }\href@noop {} {\bibfield  {journal} {\bibinfo  {journal} {Physics in Medicine and Biology}\ }\textbf {\bibinfo {volume} {68}},\ \bibinfo {pages} {055006} (\bibinfo {year} {2023})}\BibitemShut {NoStop}%
\bibitem [{\citenamefont {Hill}\ \emph {et~al.}(2002)\citenamefont {Hill}, \citenamefont {Stevens}, \citenamefont {Marsden}, \citenamefont {Allott}, \citenamefont {Turcu},\ and\ \citenamefont {Goodhead}}]{hill2002increased}%
  \BibitemOpen
  \bibfield  {author} {\bibinfo {author} {\bibfnamefont {M.}~\bibnamefont {Hill}}, \bibinfo {author} {\bibfnamefont {D.}~\bibnamefont {Stevens}}, \bibinfo {author} {\bibfnamefont {S.}~\bibnamefont {Marsden}}, \bibinfo {author} {\bibfnamefont {R.}~\bibnamefont {Allott}}, \bibinfo {author} {\bibfnamefont {I.}~\bibnamefont {Turcu}},\ and\ \bibinfo {author} {\bibfnamefont {D.}~\bibnamefont {Goodhead}},\ }\bibfield  {title} {\bibinfo {title} {Is the increased relative biological effectiveness of high let particles due to spatial or temporal effects? characterization and oer in v79-4 cells},\ }\href@noop {} {\bibfield  {journal} {\bibinfo  {journal} {Physics in Medicine \& Biology}\ }\textbf {\bibinfo {volume} {47}},\ \bibinfo {pages} {3543} (\bibinfo {year} {2002})}\BibitemShut {NoStop}%
\bibitem [{\citenamefont {McAnespie}\ \emph {et~al.}(2023)\citenamefont {McAnespie}, \citenamefont {Chaudhary}, \citenamefont {Calvin}, \citenamefont {Streeter}, \citenamefont {Nersysian}, \citenamefont {McMahon}, \citenamefont {Prise},\ and\ \citenamefont {Sarri}}]{mcanespie2023ijrobp}%
  \BibitemOpen
  \bibfield  {author} {\bibinfo {author} {\bibfnamefont {C.~A.}\ \bibnamefont {McAnespie}}, \bibinfo {author} {\bibfnamefont {P.}~\bibnamefont {Chaudhary}}, \bibinfo {author} {\bibfnamefont {L.}~\bibnamefont {Calvin}}, \bibinfo {author} {\bibfnamefont {M.~J.~V.}\ \bibnamefont {Streeter}}, \bibinfo {author} {\bibfnamefont {G.}~\bibnamefont {Nersysian}}, \bibinfo {author} {\bibfnamefont {S.~J.}\ \bibnamefont {McMahon}}, \bibinfo {author} {\bibfnamefont {K.~M.}\ \bibnamefont {Prise}},\ and\ \bibinfo {author} {\bibfnamefont {G.}~\bibnamefont {Sarri}},\ }\bibfield  {title} {\bibinfo {title} {Response of cancer stem cells and human skin fibroblasts to picosecond-scale electron irradiation at $10^{10}$ to $10^{11}$ gy/s},\ }\href {https://doi.org/10.1016/j.ijrobp.2023.10.024} {\bibfield  {journal} {\bibinfo  {journal} {Int. J. Radiat. Oncol. Biol. Phys.}\ } (\bibinfo {year} {2023})}\BibitemShut {NoStop}%
\bibitem [{\citenamefont {Beyreuther}\ \emph {et~al.}(2015)\citenamefont {Beyreuther}, \citenamefont {Karsch}, \citenamefont {Laschinsky}, \citenamefont {Le{\ss}mann}, \citenamefont {Naumburger}, \citenamefont {Oppelt}, \citenamefont {Richter}, \citenamefont {Sch{\"u}rer}, \citenamefont {Woithe},\ and\ \citenamefont {Pawelke}}]{beyreuther2015radiobiological}%
  \BibitemOpen
  \bibfield  {author} {\bibinfo {author} {\bibfnamefont {E.}~\bibnamefont {Beyreuther}}, \bibinfo {author} {\bibfnamefont {L.}~\bibnamefont {Karsch}}, \bibinfo {author} {\bibfnamefont {L.}~\bibnamefont {Laschinsky}}, \bibinfo {author} {\bibfnamefont {E.}~\bibnamefont {Le{\ss}mann}}, \bibinfo {author} {\bibfnamefont {D.}~\bibnamefont {Naumburger}}, \bibinfo {author} {\bibfnamefont {M.}~\bibnamefont {Oppelt}}, \bibinfo {author} {\bibfnamefont {C.}~\bibnamefont {Richter}}, \bibinfo {author} {\bibfnamefont {M.}~\bibnamefont {Sch{\"u}rer}}, \bibinfo {author} {\bibfnamefont {J.}~\bibnamefont {Woithe}},\ and\ \bibinfo {author} {\bibfnamefont {J.}~\bibnamefont {Pawelke}},\ }\bibfield  {title} {\bibinfo {title} {Radiobiological response to ultra-short pulsed megavoltage electron beams of ultra-high pulse dose rate},\ }\href@noop {} {\bibfield  {journal} {\bibinfo  {journal} {International Journal of Radiation Biology}\ }\textbf {\bibinfo {volume} {91}},\ \bibinfo {pages} {643} (\bibinfo {year} {2015})}\BibitemShut
  {NoStop}%
\bibitem [{\citenamefont {Oppelt}\ \emph {et~al.}(2015)\citenamefont {Oppelt}, \citenamefont {Baumann}, \citenamefont {Bergmann}, \citenamefont {Beyreuther}, \citenamefont {Br{\"u}chner}, \citenamefont {Hartmann}, \citenamefont {Karsch}, \citenamefont {Krause}, \citenamefont {Laschinsky}, \citenamefont {Le{\ss}mann} \emph {et~al.}}]{oppelt2015comparison}%
  \BibitemOpen
  \bibfield  {author} {\bibinfo {author} {\bibfnamefont {M.}~\bibnamefont {Oppelt}}, \bibinfo {author} {\bibfnamefont {M.}~\bibnamefont {Baumann}}, \bibinfo {author} {\bibfnamefont {R.}~\bibnamefont {Bergmann}}, \bibinfo {author} {\bibfnamefont {E.}~\bibnamefont {Beyreuther}}, \bibinfo {author} {\bibfnamefont {K.}~\bibnamefont {Br{\"u}chner}}, \bibinfo {author} {\bibfnamefont {J.}~\bibnamefont {Hartmann}}, \bibinfo {author} {\bibfnamefont {L.}~\bibnamefont {Karsch}}, \bibinfo {author} {\bibfnamefont {M.}~\bibnamefont {Krause}}, \bibinfo {author} {\bibfnamefont {L.}~\bibnamefont {Laschinsky}}, \bibinfo {author} {\bibfnamefont {E.}~\bibnamefont {Le{\ss}mann}}, \emph {et~al.},\ }\bibfield  {title} {\bibinfo {title} {Comparison study of in vivo dose response to laser-driven versus conventional electron beam},\ }\href@noop {} {\bibfield  {journal} {\bibinfo  {journal} {Radiation and environmental biophysics}\ }\textbf {\bibinfo {volume} {54}},\ \bibinfo {pages} {155} (\bibinfo {year} {2015})}\BibitemShut {NoStop}%
\bibitem [{\citenamefont {Acharya}\ \emph {et~al.}(2011)\citenamefont {Acharya}, \citenamefont {Bhat}, \citenamefont {Joseph}, \citenamefont {Sanjeev}, \citenamefont {Sreedevi},\ and\ \citenamefont {Narayana}}]{acharya2011dose}%
  \BibitemOpen
  \bibfield  {author} {\bibinfo {author} {\bibfnamefont {S.}~\bibnamefont {Acharya}}, \bibinfo {author} {\bibfnamefont {N.}~\bibnamefont {Bhat}}, \bibinfo {author} {\bibfnamefont {P.}~\bibnamefont {Joseph}}, \bibinfo {author} {\bibfnamefont {G.}~\bibnamefont {Sanjeev}}, \bibinfo {author} {\bibfnamefont {B.}~\bibnamefont {Sreedevi}},\ and\ \bibinfo {author} {\bibfnamefont {Y.}~\bibnamefont {Narayana}},\ }\bibfield  {title} {\bibinfo {title} {Dose rate effect on micronuclei induction in human blood lymphocytes exposed to single pulse and multiple pulses of electrons},\ }\href@noop {} {\bibfield  {journal} {\bibinfo  {journal} {Radiation and environmental biophysics}\ }\textbf {\bibinfo {volume} {50}},\ \bibinfo {pages} {253} (\bibinfo {year} {2011})}\BibitemShut {NoStop}%
\bibitem [{\citenamefont {Nias}\ \emph {et~al.}(1970)\citenamefont {Nias}, \citenamefont {Swallow}, \citenamefont {Keene},\ and\ \citenamefont {Hodgson}}]{Nias}%
  \BibitemOpen
  \bibfield  {author} {\bibinfo {author} {\bibfnamefont {A.~H.~W.}\ \bibnamefont {Nias}}, \bibinfo {author} {\bibfnamefont {A.~J.}\ \bibnamefont {Swallow}}, \bibinfo {author} {\bibfnamefont {J.~P.}\ \bibnamefont {Keene}},\ and\ \bibinfo {author} {\bibfnamefont {B.~W.}\ \bibnamefont {Hodgson}},\ }\bibfield  {title} {\bibinfo {title} {Survival of hela cells from 10 nanosecond pulses of electrons},\ }\href@noop {} {\bibfield  {journal} {\bibinfo  {journal} {Int. J. Radiat. Oncol. Biol. Phys.}\ }\textbf {\bibinfo {volume} {17}},\ \bibinfo {pages} {595} (\bibinfo {year} {1970})}\BibitemShut {NoStop}%
\bibitem [{\citenamefont {Margarone}\ \emph {et~al.}(2018)\citenamefont {Margarone} \emph {et~al.}}]{margarone2018}%
  \BibitemOpen
  \bibfield  {author} {\bibinfo {author} {\bibfnamefont {D.}~\bibnamefont {Margarone}} \emph {et~al.},\ }\bibfield  {title} {\bibinfo {title} {Elimaia: A laser-driven ion accelerator for multidisciplinary applications},\ }\href@noop {} {\bibfield  {journal} {\bibinfo  {journal} {Quantum beam science}\ }\textbf {\bibinfo {volume} {2}},\ \bibinfo {pages} {8} (\bibinfo {year} {2018})}\BibitemShut {NoStop}%
\bibitem [{\citenamefont {Hanton}\ \emph {et~al.}(2019)\citenamefont {Hanton}, \citenamefont {Chaudhary}, \citenamefont {Doria}, \citenamefont {Gwynne}, \citenamefont {Maiorino}, \citenamefont {Scullion}, \citenamefont {Ahmed}, \citenamefont {Marshall}, \citenamefont {Naughton}, \citenamefont {Romagnani} \emph {et~al.}}]{hanton2019dna}%
  \BibitemOpen
  \bibfield  {author} {\bibinfo {author} {\bibfnamefont {F.}~\bibnamefont {Hanton}}, \bibinfo {author} {\bibfnamefont {P.}~\bibnamefont {Chaudhary}}, \bibinfo {author} {\bibfnamefont {D.}~\bibnamefont {Doria}}, \bibinfo {author} {\bibfnamefont {D.}~\bibnamefont {Gwynne}}, \bibinfo {author} {\bibfnamefont {C.}~\bibnamefont {Maiorino}}, \bibinfo {author} {\bibfnamefont {C.}~\bibnamefont {Scullion}}, \bibinfo {author} {\bibfnamefont {H.}~\bibnamefont {Ahmed}}, \bibinfo {author} {\bibfnamefont {T.}~\bibnamefont {Marshall}}, \bibinfo {author} {\bibfnamefont {K.}~\bibnamefont {Naughton}}, \bibinfo {author} {\bibfnamefont {L.}~\bibnamefont {Romagnani}}, \emph {et~al.},\ }\bibfield  {title} {\bibinfo {title} {Dna dsb repair dynamics following irradiation with laser-driven protons at ultra-high dose rates},\ }\href@noop {} {\bibfield  {journal} {\bibinfo  {journal} {Scientific Reports}\ }\textbf {\bibinfo {volume} {9}},\ \bibinfo {pages} {4471} (\bibinfo {year} {2019})}\BibitemShut {NoStop}%
\bibitem [{\citenamefont {Doria}\ \emph {et~al.}(2012)\citenamefont {Doria}, \citenamefont {Kakolee}, \citenamefont {Kar}, \citenamefont {Litt}, \citenamefont {Fiorini}, \citenamefont {Ahmed}, \citenamefont {Green}, \citenamefont {Jeynes}, \citenamefont {Kavanagh}, \citenamefont {Kirby} \emph {et~al.}}]{doria2012biological}%
  \BibitemOpen
  \bibfield  {author} {\bibinfo {author} {\bibfnamefont {D.}~\bibnamefont {Doria}}, \bibinfo {author} {\bibfnamefont {K.}~\bibnamefont {Kakolee}}, \bibinfo {author} {\bibfnamefont {S.}~\bibnamefont {Kar}}, \bibinfo {author} {\bibfnamefont {S.}~\bibnamefont {Litt}}, \bibinfo {author} {\bibfnamefont {F.}~\bibnamefont {Fiorini}}, \bibinfo {author} {\bibfnamefont {H.}~\bibnamefont {Ahmed}}, \bibinfo {author} {\bibfnamefont {S.}~\bibnamefont {Green}}, \bibinfo {author} {\bibfnamefont {J.}~\bibnamefont {Jeynes}}, \bibinfo {author} {\bibfnamefont {J.}~\bibnamefont {Kavanagh}}, \bibinfo {author} {\bibfnamefont {D.}~\bibnamefont {Kirby}}, \emph {et~al.},\ }\bibfield  {title} {\bibinfo {title} {Biological effectiveness on live cells of laser driven protons at dose rates exceeding $10^9$ gy/s},\ }\href@noop {} {\bibfield  {journal} {\bibinfo  {journal} {AIP Advances}\ }\textbf {\bibinfo {volume} {2}},\ \bibinfo {pages} {011209} (\bibinfo {year} {2012})}\BibitemShut {NoStop}%
\bibitem [{\citenamefont {Kroll}\ \emph {et~al.}(2022)\citenamefont {Kroll} \emph {et~al.}}]{kroll_2022}%
  \BibitemOpen
  \bibfield  {author} {\bibinfo {author} {\bibfnamefont {F.}~\bibnamefont {Kroll}} \emph {et~al.},\ }\bibfield  {title} {\bibinfo {title} {Tumour irradiation in mice with a laser-accelerated proton beam},\ }\href@noop {} {\bibfield  {journal} {\bibinfo  {journal} {Nature Physics}\ }\textbf {\bibinfo {volume} {18}},\ \bibinfo {pages} {316} (\bibinfo {year} {2022})}\BibitemShut {NoStop}%
\bibitem [{\citenamefont {Esarey}\ \emph {et~al.}(2009)\citenamefont {Esarey}, \citenamefont {Schroeder},\ and\ \citenamefont {Leemans}}]{esarey2009physics}%
  \BibitemOpen
  \bibfield  {author} {\bibinfo {author} {\bibfnamefont {E.}~\bibnamefont {Esarey}}, \bibinfo {author} {\bibfnamefont {C.~B.}\ \bibnamefont {Schroeder}},\ and\ \bibinfo {author} {\bibfnamefont {W.~P.}\ \bibnamefont {Leemans}},\ }\bibfield  {title} {\bibinfo {title} {Physics of laser-driven plasma-based electron accelerators},\ }\href@noop {} {\bibfield  {journal} {\bibinfo  {journal} {Reviews of modern physics}\ }\textbf {\bibinfo {volume} {81}},\ \bibinfo {pages} {1229} (\bibinfo {year} {2009})}\BibitemShut {NoStop}%
\bibitem [{\citenamefont {Poder}\ \emph {et~al.}(2024)\citenamefont {Poder} \emph {et~al.}}]{poder_2024}%
  \BibitemOpen
  \bibfield  {author} {\bibinfo {author} {\bibfnamefont {K.}~\bibnamefont {Poder}} \emph {et~al.},\ }\bibfield  {title} {\bibinfo {title} {Multi-gev electron acceleration in wakefields strongly driven by oversized laser spots},\ }\href@noop {} {\bibfield  {journal} {\bibinfo  {journal} {Physical Review Letters}\ }\textbf {\bibinfo {volume} {132}},\ \bibinfo {pages} {195001} (\bibinfo {year} {2024})}\BibitemShut {NoStop}%
\bibitem [{\citenamefont {McAnespie}\ \emph {et~al.}(2024)\citenamefont {McAnespie}, \citenamefont {Chaudhary}, \citenamefont {Calvin}, \citenamefont {Streeter}, \citenamefont {McmMahon}, \citenamefont {Prise},\ and\ \citenamefont {Sarri}}]{McAnespie_pre}%
  \BibitemOpen
  \bibfield  {author} {\bibinfo {author} {\bibfnamefont {C.}~\bibnamefont {McAnespie}}, \bibinfo {author} {\bibfnamefont {P.}~\bibnamefont {Chaudhary}}, \bibinfo {author} {\bibfnamefont {L.}~\bibnamefont {Calvin}}, \bibinfo {author} {\bibfnamefont {M.}~\bibnamefont {Streeter}}, \bibinfo {author} {\bibfnamefont {S.}~\bibnamefont {McmMahon}}, \bibinfo {author} {\bibfnamefont {K.}~\bibnamefont {Prise}},\ and\ \bibinfo {author} {\bibfnamefont {G.}~\bibnamefont {Sarri}},\ }\bibfield  {title} {\bibinfo {title} {Laser-driven electron source suitable for single-shot gy-scale irradiation of biological cells at dose-rates exceeding $10^{10}$ gy/s.},\ }\href@noop {} {\bibfield  {journal} {\bibinfo  {journal} {Physical Review E}\ }\textbf {\bibinfo {volume} {110}},\ \bibinfo {pages} {035204} (\bibinfo {year} {2024})}\BibitemShut {NoStop}%
\bibitem [{\citenamefont {Murshed}(2019)}]{Murshed}%
  \BibitemOpen
  \bibfield  {author} {\bibinfo {author} {\bibfnamefont {H.}~\bibnamefont {Murshed}},\ }\bibfield  {title} {\bibinfo {title} {Fundamentals of radiation oncology},\ }\href@noop {} {\bibfield  {journal} {\bibinfo  {journal} {Elsevier}\ } (\bibinfo {year} {2019})}\BibitemShut {NoStop}%
\bibitem [{\citenamefont {Delorme}\ \emph {et~al.}(2021)\citenamefont {Delorme} \emph {et~al.}}]{Delorme}%
  \BibitemOpen
  \bibfield  {author} {\bibinfo {author} {\bibfnamefont {R.}~\bibnamefont {Delorme}} \emph {et~al.},\ }\bibfield  {title} {\bibinfo {title} {First theoretical determination of relative biological effectiveness of very high energy electrons},\ }\href@noop {} {\bibfield  {journal} {\bibinfo  {journal} {Scientific Reports}\ }\textbf {\bibinfo {volume} {11}},\ \bibinfo {pages} {11242} (\bibinfo {year} {2021})}\BibitemShut {NoStop}%
\bibitem [{\citenamefont {Pak}\ \emph {et~al.}(2010)\citenamefont {Pak}, \citenamefont {Marsh}, \citenamefont {Martins}, \citenamefont {Lu}, \citenamefont {Mori},\ and\ \citenamefont {Joshi}}]{pak2010injection}%
  \BibitemOpen
  \bibfield  {author} {\bibinfo {author} {\bibfnamefont {A.}~\bibnamefont {Pak}}, \bibinfo {author} {\bibfnamefont {K.}~\bibnamefont {Marsh}}, \bibinfo {author} {\bibfnamefont {S.}~\bibnamefont {Martins}}, \bibinfo {author} {\bibfnamefont {W.}~\bibnamefont {Lu}}, \bibinfo {author} {\bibfnamefont {W.}~\bibnamefont {Mori}},\ and\ \bibinfo {author} {\bibfnamefont {C.}~\bibnamefont {Joshi}},\ }\bibfield  {title} {\bibinfo {title} {Injection and trapping of tunnel-ionized electrons into laser-produced wakes},\ }\href@noop {} {\bibfield  {journal} {\bibinfo  {journal} {Physical Review Letters}\ }\textbf {\bibinfo {volume} {104}},\ \bibinfo {pages} {025003} (\bibinfo {year} {2010})}\BibitemShut {NoStop}%
\bibitem [{\citenamefont {Mirzaie}\ \emph {et~al.}(2015)\citenamefont {Mirzaie}, \citenamefont {Li}, \citenamefont {Zeng}, \citenamefont {Hafz}, \citenamefont {Chen}, \citenamefont {Li}, \citenamefont {Zhu}, \citenamefont {Liao}, \citenamefont {Sokollik}, \citenamefont {Liu} \emph {et~al.}}]{mirzaie2015demonstration}%
  \BibitemOpen
  \bibfield  {author} {\bibinfo {author} {\bibfnamefont {M.}~\bibnamefont {Mirzaie}}, \bibinfo {author} {\bibfnamefont {S.}~\bibnamefont {Li}}, \bibinfo {author} {\bibfnamefont {M.}~\bibnamefont {Zeng}}, \bibinfo {author} {\bibfnamefont {N.}~\bibnamefont {Hafz}}, \bibinfo {author} {\bibfnamefont {M.}~\bibnamefont {Chen}}, \bibinfo {author} {\bibfnamefont {G.}~\bibnamefont {Li}}, \bibinfo {author} {\bibfnamefont {Q.}~\bibnamefont {Zhu}}, \bibinfo {author} {\bibfnamefont {H.}~\bibnamefont {Liao}}, \bibinfo {author} {\bibfnamefont {T.}~\bibnamefont {Sokollik}}, \bibinfo {author} {\bibfnamefont {F.}~\bibnamefont {Liu}}, \emph {et~al.},\ }\bibfield  {title} {\bibinfo {title} {Demonstration of self-truncated ionization injection for gev electron beams},\ }\href@noop {} {\bibfield  {journal} {\bibinfo  {journal} {Scientific reports}\ }\textbf {\bibinfo {volume} {5}},\ \bibinfo {pages} {1} (\bibinfo {year} {2015})}\BibitemShut {NoStop}%
\bibitem [{\citenamefont {Lu}\ \emph {et~al.}(2007)\citenamefont {Lu} \emph {et~al.}}]{lu_2007}%
  \BibitemOpen
  \bibfield  {author} {\bibinfo {author} {\bibfnamefont {W.}~\bibnamefont {Lu}} \emph {et~al.},\ }\bibfield  {title} {\bibinfo {title} {Generating multi-gev electron bunches using single stage laser wakefield acceleration in a 3d nonlinear regime},\ }\href@noop {} {\bibfield  {journal} {\bibinfo  {journal} {Physical Review Special Topics - Accelerators and Beams}\ }\textbf {\bibinfo {volume} {10}},\ \bibinfo {pages} {061301} (\bibinfo {year} {2007})}\BibitemShut {NoStop}%
\bibitem [{\citenamefont {Glinec}\ \emph {et~al.}(2006)\citenamefont {Glinec}, \citenamefont {Faure}, \citenamefont {Guemnie-Tafo}, \citenamefont {Malka}, \citenamefont {Monard}, \citenamefont {Larbre}, \citenamefont {De~Waele}, \citenamefont {Marignier},\ and\ \citenamefont {Mostafavi}}]{glinec2006absolute}%
  \BibitemOpen
  \bibfield  {author} {\bibinfo {author} {\bibfnamefont {Y.}~\bibnamefont {Glinec}}, \bibinfo {author} {\bibfnamefont {J.}~\bibnamefont {Faure}}, \bibinfo {author} {\bibfnamefont {A.}~\bibnamefont {Guemnie-Tafo}}, \bibinfo {author} {\bibfnamefont {V.}~\bibnamefont {Malka}}, \bibinfo {author} {\bibfnamefont {H.}~\bibnamefont {Monard}}, \bibinfo {author} {\bibfnamefont {J.}~\bibnamefont {Larbre}}, \bibinfo {author} {\bibfnamefont {V.}~\bibnamefont {De~Waele}}, \bibinfo {author} {\bibfnamefont {J.}~\bibnamefont {Marignier}},\ and\ \bibinfo {author} {\bibfnamefont {M.}~\bibnamefont {Mostafavi}},\ }\bibfield  {title} {\bibinfo {title} {Absolute calibration for a broad range single shot electron spectrometer},\ }\href@noop {} {\bibfield  {journal} {\bibinfo  {journal} {Review of scientific instruments}\ }\textbf {\bibinfo {volume} {77}} (\bibinfo {year} {2006})}\BibitemShut {NoStop}%
\bibitem [{\citenamefont {Mangles}\ \emph {et~al.}(2012)\citenamefont {Mangles} \emph {et~al.}}]{Mangles_2012}%
  \BibitemOpen
  \bibfield  {author} {\bibinfo {author} {\bibfnamefont {S.}~\bibnamefont {Mangles}} \emph {et~al.},\ }\bibfield  {title} {\bibinfo {title} {Self-injection threshold in self-guided laser wakefield accelerators},\ }\href@noop {} {\bibfield  {journal} {\bibinfo  {journal} {Physical Review Special Topics - Accelerators and Beams}\ }\textbf {\bibinfo {volume} {15}},\ \bibinfo {pages} {011302} (\bibinfo {year} {2012})}\BibitemShut {NoStop}%
\bibitem [{\citenamefont {Lehe}\ \emph {et~al.}(2016)\citenamefont {Lehe} \emph {et~al.}}]{FBPIC}%
  \BibitemOpen
  \bibfield  {author} {\bibinfo {author} {\bibfnamefont {R.}~\bibnamefont {Lehe}} \emph {et~al.},\ }\bibfield  {title} {\bibinfo {title} {A spectral, quasi-cylindrical and dispersion-free particle-in-cell algorithm},\ }\href@noop {} {\bibfield  {journal} {\bibinfo  {journal} {Comput. Phys. Commun.}\ }\textbf {\bibinfo {volume} {203}},\ \bibinfo {pages} {66} (\bibinfo {year} {2016})}\BibitemShut {NoStop}%
\bibitem [{\citenamefont {Marroquin}\ \emph {et~al.}(2016)\citenamefont {Marroquin}, \citenamefont {Herrera~Gonzalez}, \citenamefont {Camacho~Lopez}, \citenamefont {Barajas},\ and\ \citenamefont {Garc{\'\i}a-Gardu{\~n}o}}]{marroquin2016evaluation}%
  \BibitemOpen
  \bibfield  {author} {\bibinfo {author} {\bibfnamefont {E.~Y.~L.}\ \bibnamefont {Marroquin}}, \bibinfo {author} {\bibfnamefont {J.~A.}\ \bibnamefont {Herrera~Gonzalez}}, \bibinfo {author} {\bibfnamefont {M.~A.}\ \bibnamefont {Camacho~Lopez}}, \bibinfo {author} {\bibfnamefont {J.~E.~V.}\ \bibnamefont {Barajas}},\ and\ \bibinfo {author} {\bibfnamefont {O.~A.}\ \bibnamefont {Garc{\'\i}a-Gardu{\~n}o}},\ }\bibfield  {title} {\bibinfo {title} {Evaluation of the uncertainty in an ebt3 film dosimetry system utilizing net optical density},\ }\href@noop {} {\bibfield  {journal} {\bibinfo  {journal} {Journal of applied clinical medical physics}\ }\textbf {\bibinfo {volume} {17}},\ \bibinfo {pages} {466} (\bibinfo {year} {2016})}\BibitemShut {NoStop}%
\bibitem [{\citenamefont {Perl}\ \emph {et~al.}(2012)\citenamefont {Perl}, \citenamefont {Shin}, \citenamefont {Sch{\"u}mann}, \citenamefont {Faddegon},\ and\ \citenamefont {Paganetti}}]{perl2012topas}%
  \BibitemOpen
  \bibfield  {author} {\bibinfo {author} {\bibfnamefont {J.}~\bibnamefont {Perl}}, \bibinfo {author} {\bibfnamefont {J.}~\bibnamefont {Shin}}, \bibinfo {author} {\bibfnamefont {J.}~\bibnamefont {Sch{\"u}mann}}, \bibinfo {author} {\bibfnamefont {B.}~\bibnamefont {Faddegon}},\ and\ \bibinfo {author} {\bibfnamefont {H.}~\bibnamefont {Paganetti}},\ }\bibfield  {title} {\bibinfo {title} {Topas: an innovative proton monte carlo platform for research and clinical applications},\ }\href@noop {} {\bibfield  {journal} {\bibinfo  {journal} {Medical physics}\ }\textbf {\bibinfo {volume} {39}},\ \bibinfo {pages} {6818} (\bibinfo {year} {2012})}\BibitemShut {NoStop}%
\bibitem [{\citenamefont {Agostinelli}\ \emph {et~al.}(2003)\citenamefont {Agostinelli}, \citenamefont {Allison}, \citenamefont {Amako}, \citenamefont {Apostolakis}, \citenamefont {Araujo}, \citenamefont {Arce}, \citenamefont {Asai}, \citenamefont {Axen}, \citenamefont {Banerjee}, \citenamefont {Barrand} \emph {et~al.}}]{agostinelli2003geant4}%
  \BibitemOpen
  \bibfield  {author} {\bibinfo {author} {\bibfnamefont {S.}~\bibnamefont {Agostinelli}}, \bibinfo {author} {\bibfnamefont {J.}~\bibnamefont {Allison}}, \bibinfo {author} {\bibfnamefont {K.~a.}\ \bibnamefont {Amako}}, \bibinfo {author} {\bibfnamefont {J.}~\bibnamefont {Apostolakis}}, \bibinfo {author} {\bibfnamefont {H.}~\bibnamefont {Araujo}}, \bibinfo {author} {\bibfnamefont {P.}~\bibnamefont {Arce}}, \bibinfo {author} {\bibfnamefont {M.}~\bibnamefont {Asai}}, \bibinfo {author} {\bibfnamefont {D.}~\bibnamefont {Axen}}, \bibinfo {author} {\bibfnamefont {S.}~\bibnamefont {Banerjee}}, \bibinfo {author} {\bibfnamefont {G.}~\bibnamefont {Barrand}}, \emph {et~al.},\ }\bibfield  {title} {\bibinfo {title} {Geant4—a simulation toolkit},\ }\href@noop {} {\bibfield  {journal} {\bibinfo  {journal} {Nuclear instruments and methods in physics research section A: Accelerators, Spectrometers, Detectors and Associated Equipment}\ }\textbf {\bibinfo {volume} {506}},\ \bibinfo {pages} {250} (\bibinfo {year}
  {2003})}\BibitemShut {NoStop}%
\bibitem [{\citenamefont {Allison}\ \emph {et~al.}(2016)\citenamefont {Allison}, \citenamefont {Amako}, \citenamefont {Apostolakis}, \citenamefont {Arce}, \citenamefont {Asai}, \citenamefont {Aso}, \citenamefont {Bagli}, \citenamefont {Bagulya}, \citenamefont {Banerjee}, \citenamefont {Barrand} \emph {et~al.}}]{allison2016recent}%
  \BibitemOpen
  \bibfield  {author} {\bibinfo {author} {\bibfnamefont {J.}~\bibnamefont {Allison}}, \bibinfo {author} {\bibfnamefont {K.}~\bibnamefont {Amako}}, \bibinfo {author} {\bibfnamefont {J.}~\bibnamefont {Apostolakis}}, \bibinfo {author} {\bibfnamefont {P.}~\bibnamefont {Arce}}, \bibinfo {author} {\bibfnamefont {M.}~\bibnamefont {Asai}}, \bibinfo {author} {\bibfnamefont {T.}~\bibnamefont {Aso}}, \bibinfo {author} {\bibfnamefont {E.}~\bibnamefont {Bagli}}, \bibinfo {author} {\bibfnamefont {A.}~\bibnamefont {Bagulya}}, \bibinfo {author} {\bibfnamefont {S.}~\bibnamefont {Banerjee}}, \bibinfo {author} {\bibfnamefont {G.}~\bibnamefont {Barrand}}, \emph {et~al.},\ }\bibfield  {title} {\bibinfo {title} {Recent developments in geant4},\ }\href@noop {} {\bibfield  {journal} {\bibinfo  {journal} {Nuclear instruments and methods in physics research section A: Accelerators, Spectrometers, Detectors and Associated Equipment}\ }\textbf {\bibinfo {volume} {835}},\ \bibinfo {pages} {186} (\bibinfo {year} {2016})}\BibitemShut
  {NoStop}%
\bibitem [{\citenamefont {Maier}\ \emph {et~al.}(2020)\citenamefont {Maier} \emph {et~al.}}]{Maier}%
  \BibitemOpen
  \bibfield  {author} {\bibinfo {author} {\bibfnamefont {A.}~\bibnamefont {Maier}} \emph {et~al.},\ }\bibfield  {title} {\bibinfo {title} {Decoding sources of energy variability in a laser-plasma accelerator},\ }\href@noop {} {\bibfield  {journal} {\bibinfo  {journal} {Phys. Rev. X}\ }\textbf {\bibinfo {volume} {10}},\ \bibinfo {pages} {031039} (\bibinfo {year} {2020})}\BibitemShut {NoStop}%
\bibitem [{\citenamefont {Chaudhary}\ \emph {et~al.}(2022)\citenamefont {Chaudhary}, \citenamefont {Gwynne}, \citenamefont {Odlozilik}, \citenamefont {McMurray}, \citenamefont {Milluzzo}, \citenamefont {Maiorino}, \citenamefont {Doria}, \citenamefont {Ahmed}, \citenamefont {Romagnani}, \citenamefont {Alejo}, \citenamefont {A~Padda}, \citenamefont {Green}, \citenamefont {Carroll}, \citenamefont {Booth}, \citenamefont {McKenna}, \citenamefont {Kar}, \citenamefont {Petringa}, \citenamefont {Catalano}, \citenamefont {Cammarata}, \citenamefont {Cirrone}, \citenamefont {McMahon}, \citenamefont {Prise},\ and\ \citenamefont {K~Borghesi}}]{Chaudhary2022}%
  \BibitemOpen
  \bibfield  {author} {\bibinfo {author} {\bibfnamefont {P.}~\bibnamefont {Chaudhary}}, \bibinfo {author} {\bibfnamefont {D.}~\bibnamefont {Gwynne}}, \bibinfo {author} {\bibfnamefont {B.}~\bibnamefont {Odlozilik}}, \bibinfo {author} {\bibfnamefont {A.}~\bibnamefont {McMurray}}, \bibinfo {author} {\bibfnamefont {G.}~\bibnamefont {Milluzzo}}, \bibinfo {author} {\bibfnamefont {C.}~\bibnamefont {Maiorino}}, \bibinfo {author} {\bibfnamefont {D.}~\bibnamefont {Doria}}, \bibinfo {author} {\bibfnamefont {H.}~\bibnamefont {Ahmed}}, \bibinfo {author} {\bibfnamefont {L.}~\bibnamefont {Romagnani}}, \bibinfo {author} {\bibfnamefont {A.}~\bibnamefont {Alejo}}, \bibinfo {author} {\bibfnamefont {H.}~\bibnamefont {A~Padda}}, \bibinfo {author} {\bibfnamefont {J.}~\bibnamefont {Green}}, \bibinfo {author} {\bibfnamefont {D.}~\bibnamefont {Carroll}}, \bibinfo {author} {\bibfnamefont {N.}~\bibnamefont {Booth}}, \bibinfo {author} {\bibfnamefont {P.}~\bibnamefont {McKenna}}, \bibinfo {author} {\bibfnamefont {S.}~\bibnamefont {Kar}},
  \bibinfo {author} {\bibfnamefont {G.}~\bibnamefont {Petringa}}, \bibinfo {author} {\bibfnamefont {R.}~\bibnamefont {Catalano}}, \bibinfo {author} {\bibfnamefont {F.}~\bibnamefont {Cammarata}}, \bibinfo {author} {\bibfnamefont {P.}~\bibnamefont {Cirrone}}, \bibinfo {author} {\bibfnamefont {S.}~\bibnamefont {McMahon}}, \bibinfo {author} {\bibnamefont {Prise}},\ and\ \bibinfo {author} {\bibfnamefont {M.}~\bibnamefont {K~Borghesi}},\ }\bibfield  {title} {\bibinfo {title} {Development of a portable hypoxia chamber for ultra-high dose rate laser-driven proton radiobiology applications},\ }\href@noop {} {\bibfield  {journal} {\bibinfo  {journal} {Radiation Oncology}\ }\textbf {\bibinfo {volume} {17}},\ \bibinfo {pages} {77} (\bibinfo {year} {2022})}\BibitemShut {NoStop}%
\bibitem [{\citenamefont {Feoktistova}\ \emph {et~al.}(2016)\citenamefont {Feoktistova}, \citenamefont {Geserick},\ and\ \citenamefont {Leverkus}}]{feoktistova2016crystal}%
  \BibitemOpen
  \bibfield  {author} {\bibinfo {author} {\bibfnamefont {M.}~\bibnamefont {Feoktistova}}, \bibinfo {author} {\bibfnamefont {P.}~\bibnamefont {Geserick}},\ and\ \bibinfo {author} {\bibfnamefont {M.}~\bibnamefont {Leverkus}},\ }\bibfield  {title} {\bibinfo {title} {Crystal violet assay for determining viability of cultured cells},\ }\href@noop {} {\bibfield  {journal} {\bibinfo  {journal} {Cold Spring Harbor Protocols}\ }\textbf {\bibinfo {volume} {2016}},\ \bibinfo {pages} {pdb} (\bibinfo {year} {2016})}\BibitemShut {NoStop}%
\bibitem [{sup()}]{suppl}%
  \BibitemOpen
  \bibfield  {title} {\bibinfo {title} {Supplementary material showing full survival curves (fig.1) and exemplary split-channel images of foci measurements for all irradiation conditions and for both cell lines (figs. 2 - 9)},\ }\href@noop {} {\bibinfo  {journal} {web link:...}\ }\BibitemShut {NoStop}%
\bibitem [{\citenamefont {McMahon}(2018)}]{mcmahon2018linear}%
  \BibitemOpen
\bibfield  {journal} {  }\bibfield  {author} {\bibinfo {author} {\bibfnamefont {S.~J.}\ \bibnamefont {McMahon}},\ }\bibfield  {title} {\bibinfo {title} {The linear quadratic model: usage, interpretation and challenges},\ }\href@noop {} {\bibfield  {journal} {\bibinfo  {journal} {Physics in Medicine \& Biology}\ }\textbf {\bibinfo {volume} {64}},\ \bibinfo {pages} {01TR01} (\bibinfo {year} {2018})}\BibitemShut {NoStop}%
\bibitem [{\citenamefont {Joiner}\ and\ \citenamefont {Van~der Kogel}(2009)}]{kogel}%
  \BibitemOpen
  \bibfield  {author} {\bibinfo {author} {\bibfnamefont {M.}~\bibnamefont {Joiner}}\ and\ \bibinfo {author} {\bibfnamefont {A.}~\bibnamefont {Van~der Kogel}},\ }\href@noop {} {\emph {\bibinfo {title} {Basic Clinical Radiobiology}}}\ (\bibinfo  {publisher} {Hodder Arnold},\ \bibinfo {year} {2009})\BibitemShut {NoStop}%
\bibitem [{\citenamefont {Wilson}\ \emph {et~al.}(2020)\citenamefont {Wilson}, \citenamefont {Hammond}, \citenamefont {Higgins},\ and\ \citenamefont {Petersson}}]{wilson2020ultra}%
  \BibitemOpen
  \bibfield  {author} {\bibinfo {author} {\bibfnamefont {J.~D.}\ \bibnamefont {Wilson}}, \bibinfo {author} {\bibfnamefont {E.~M.}\ \bibnamefont {Hammond}}, \bibinfo {author} {\bibfnamefont {G.~S.}\ \bibnamefont {Higgins}},\ and\ \bibinfo {author} {\bibfnamefont {K.}~\bibnamefont {Petersson}},\ }\bibfield  {title} {\bibinfo {title} {Ultra-high dose rate (flash) radiotherapy: silver bullet or fool's gold?},\ }\href@noop {} {\bibfield  {journal} {\bibinfo  {journal} {Frontiers in oncology}\ }\textbf {\bibinfo {volume} {9}},\ \bibinfo {pages} {1563} (\bibinfo {year} {2020})}\BibitemShut {NoStop}%
\bibitem [{\citenamefont {Jansen}\ \emph {et~al.}(2021)\citenamefont {Jansen}, \citenamefont {Knoll}, \citenamefont {Beyreuther}, \citenamefont {Pawelke}, \citenamefont {Skuza}, \citenamefont {Hanley}, \citenamefont {Brons}, \citenamefont {Pagliari},\ and\ \citenamefont {Seco}}]{jansen2021does}%
  \BibitemOpen
  \bibfield  {author} {\bibinfo {author} {\bibfnamefont {J.}~\bibnamefont {Jansen}}, \bibinfo {author} {\bibfnamefont {J.}~\bibnamefont {Knoll}}, \bibinfo {author} {\bibfnamefont {E.}~\bibnamefont {Beyreuther}}, \bibinfo {author} {\bibfnamefont {J.}~\bibnamefont {Pawelke}}, \bibinfo {author} {\bibfnamefont {R.}~\bibnamefont {Skuza}}, \bibinfo {author} {\bibfnamefont {R.}~\bibnamefont {Hanley}}, \bibinfo {author} {\bibfnamefont {S.}~\bibnamefont {Brons}}, \bibinfo {author} {\bibfnamefont {F.}~\bibnamefont {Pagliari}},\ and\ \bibinfo {author} {\bibfnamefont {J.}~\bibnamefont {Seco}},\ }\bibfield  {title} {\bibinfo {title} {Does flash deplete oxygen? experimental evaluation for photons, protons, and carbon ions},\ }\href@noop {} {\bibfield  {journal} {\bibinfo  {journal} {Medical physics}\ }\textbf {\bibinfo {volume} {48}},\ \bibinfo {pages} {3982} (\bibinfo {year} {2021})}\BibitemShut {NoStop}%
\bibitem [{\citenamefont {Watts}\ \emph {et~al.}(1978)\citenamefont {Watts}, \citenamefont {Maughan},\ and\ \citenamefont {Michael}}]{watts1978fast}%
  \BibitemOpen
  \bibfield  {author} {\bibinfo {author} {\bibfnamefont {M.}~\bibnamefont {Watts}}, \bibinfo {author} {\bibfnamefont {R.}~\bibnamefont {Maughan}},\ and\ \bibinfo {author} {\bibfnamefont {B.}~\bibnamefont {Michael}},\ }\bibfield  {title} {\bibinfo {title} {Fast kinetics of the oxygen effect in irradiated mammalian cells},\ }\href@noop {} {\bibfield  {journal} {\bibinfo  {journal} {International Journal of Radiation Biology and Related Studies in Physics, Chemistry and Medicine}\ }\textbf {\bibinfo {volume} {33}},\ \bibinfo {pages} {195} (\bibinfo {year} {1978})}\BibitemShut {NoStop}%
\bibitem [{\citenamefont {Roots}\ and\ \citenamefont {Shigefumi}(1975)}]{Roots1975}%
  \BibitemOpen
  \bibfield  {author} {\bibinfo {author} {\bibfnamefont {R.}~\bibnamefont {Roots}}\ and\ \bibinfo {author} {\bibfnamefont {O.}~\bibnamefont {Shigefumi}},\ }\bibfield  {title} {\bibinfo {title} {Estimation of life times and diffusion distances of radicals involved in x-ray-induced dna strand breaks or killing of mammalian},\ }\href@noop {} {\bibfield  {journal} {\bibinfo  {journal} {Radiation Research}\ }\textbf {\bibinfo {volume} {64}},\ \bibinfo {pages} {306} (\bibinfo {year} {1975})}\BibitemShut {NoStop}%
\bibitem [{\citenamefont {Hongyu}\ \emph {et~al.}(2020)\citenamefont {Hongyu} \emph {et~al.}}]{hongyu}%
  \BibitemOpen
  \bibfield  {author} {\bibinfo {author} {\bibfnamefont {Z.}~\bibnamefont {Hongyu}} \emph {et~al.},\ }\bibfield  {title} {\bibinfo {title} {A parameter sensitivity study for simulating dna damage after proton irradiation using topas-nbio},\ }\href@noop {} {\bibfield  {journal} {\bibinfo  {journal} {Phys. Med. Biol.}\ }\textbf {\bibinfo {volume} {65}},\ \bibinfo {pages} {085015} (\bibinfo {year} {2020})}\BibitemShut {NoStop}%
\end{thebibliography}%

\end{document}